\documentclass[useAMS,usenatbib]{mn2e}

\usepackage{graphicx}
\usepackage{threeparttable}
\usepackage{amsmath}

\title[Non-Gaussianity and Point Sources]{Forecasts on the
  contamination induced by unresolved point sources in primordial
  non-Gaussianity beyond Planck}
\author[A. Curto et al.]{A. Curto$^{1,}$$^{2}$
  \thanks{E-mail:curto@ifca.unican.es,acurto@mrao.cam.ac.uk} M. Tucci$^{3,}$$^{4}$
J. Gonz\'alez-Nuevo$^1$ L. Toffolatti$^{5}$ E. Mart\'{\i}nez-Gonz\'alez$^1$
\newauthor F. Arg\"ueso$^{6}$ A. Lapi$^{7,}$$^{8}$ and M. L\'opez-Caniego$^1$
\\
$^{1}$Instituto de F\'isica de Cantabria, CSIC-Universidad de Cantabria, Avda. de los Castros s/n, 39005 Santander, Spain\\
$^{2}$Astrophysics Group, Cavendish Laboratory, Madingley Road, Cambridge CB3 0H3, U.K.\\
$^{3}$LAL, Univ Paris-Sud, CNRS/IN2P3, Orsay, France\\
$^{4}$D\'epartement de Physique Th\'eorique and Center for
Astroparticle Physics, Universit\'e de Gen\`eve, \\~~24 quai Ansermet,
CH--1211 Gen\`eve 4, Switzerland \\
$^{5}$Departamento de F\'isica, Universidad de Oviedo, c/ Calvo Sotelo s/n, 33007 Oviedo, Spain\\
$^{6}$Departamento de Matem\'aticas, Universidad de Oviedo, c/ Calvo Sotelo s/n, 33007 Oviedo, Spain\\
$^{7}$Dipartimento Fisica,  Universit\`a Tor Vergata, Via Ricerca Scientifica 1, 00133 Roma, Italy \\
$^{8}$Astrophysics Sector, SISSA, Via Bonomea 265, 34136 Trieste, Italy}
\begin{document}
\date{Accepted  Received ; in original form }
%
%
\maketitle
\begin{abstract}
  In this paper we present forecasts of the contamination on different
  shapes of the primordial non-Gaussianity $f_{nl}$ parameter --
  detectable on future Cosmic Microwave Background (CMB)
  high--resolution anisotropy maps -- produced by unresolved
  extragalactic point sources at frequencies of cosmological interest
  (45--375 GHz). We consider two scenarios: an ideal (noiseless)
  mission and a possible future space-borne satellite, with
  instrumental characteristics similar to the ones proposed for the
  Cosmic Origins Explorer (COrE). The local, equilateral, orthogonal
  and flat shapes are considered in both temperature (intensity) and
  polarized emission data. The angular power spectrum and bispectrum
  of extragalactic point sources are estimated by state-of-the-art
  models of source number counts. The impact of all the most relevant
  (far--IR and radio selected) source populations on these shapes at
  COrE frequencies is studied. The results of this analysis show that
  unresolved extragalactic point sources should not induce a very
  relevant non-Gaussian signal in the frequency range 100--200\,GHz,
  thus not preventing a correct estimate of the CMB primordial
  $f_{nl}$ parameter. Polarization information allows one to
  significantly reduce the error--bars in the $f_{nl}$ parameter and
  the bias induced by unresolved sources and, hence, to widen the
  range of frequencies for $f_{nl}$ studies. On the contrary, at
  $\nu<100\,$ GHz or $\nu>225\,$GHz, important non-Gaussian deviations
  in CMB anisotropy maps are expected due to unresolved extragalactic
  sources.
\end{abstract}
\section{Introduction}
The high precision achieved during the last years to detect possible
weak non-Gaussian signals with primordial origin in the most precise
CMB anisotropy data maps (such as
WMAP\footnote{http://map.gsfc.nasa.gov/} data, the forthcoming
Planck\footnote{http://www.esa.int/planck} data and future missions)
stresses the importance of the characterization of all the possible
non-primordial contaminants. The standard models of inflation
\citep{starobinski1979,guth,albrecht,linde1982,linde1983,mukhanov1992}
predict that the primordial fluctuations, linearly imprinted in the
CMB anisotropies, are nearly compatible with a Gaussian
distribution. Any departure from Gaussianity with a primordial origin
would challenge our understanding of the physics of the early
Universe. From a practical point of view, the first approach to detect
non-Gaussian deviations is the study of the third order moments (the
so-called bispectrum) of the anisotropy map. For most of the cases
with a physical interest, this bispectrum is usually characterised by
a single amplitude, the non-linear coupling parameter $f_{nl}$
\citep[see
  e.g.][]{verde2000,komatsu2001,bartolo2004,babich2004,babich2005}. For
recent reviews of the status of the primordial non-Gaussianity
characterization see e.g.
\citet{bartolo2010,yadav2010,liguori2010,martinez2012}.

The best estimates of the $f_{nl}$ parameter can currently be obtained
using the publicly available WMAP data. Three different shapes with
relevant interest for the physics of inflation have been constrained
using these data: the local, equilateral and orthogonal $f_{nl}$
shapes. Information about the inflationary scenarios that can produce
these and other non-Gaussian shapes can be found for example in the
reviews by \citet{bartolo2004,yadav2010}. The instrumental noise
contamination and the beam size of the WMAP 61 and 94 GHz frequency
channels allow a maximum resolution of $\ell_{max}~ \sim$ 1000. The
best estimates of these shapes provided by the WMAP collaboration
\citep{komatsu2011} are $ \hat{f}^{loc}_{nl}= 32 \pm 21$,
$\hat{f}^{eq}_{nl}= 26 \pm 140$ and $\hat{f}^{ort}_{nl}= -202 \pm 104$
(68\% CL) for the local, equilateral and the orthogonal shapes
respectively \citep[see e.g.][for independent analyses with similar
  results]{curto2011b,curto2012}.

Unresolved point sources introduce non-Gaussian fluctuations in CMB
anisotropies at a detectable level. In the past, their
``contamination'' to the CMB angular bispectrum has been studied in
detail by many authors
\citep[e.g.][]{refregier2000,komatsu2001,argueso2003,gonzaleznuevo2005,argueso2006,babich2008,serra2008}. As
an example, the contamination of the $f_{nl}$ parameter due to point
sources in WMAP data has been studied by \citet{komatsu2009}
considering the simple constant-flux model to simulate the radio
sources. They found that the bias introduced in the local and the
equilateral shapes is $ \Delta f^{(loc)}_{nl} \sim 2$, $\Delta
f^{(eq)}_{nl}\sim 22$. Taking into account the source number counts
model by \citet{zotti2005}, \citet{curto2011b} estimated that the bias
in $f_{nl}$ is $ \Delta f^{(loc)}_{nl}\sim 2.5$, $\Delta
f^{(eq)}_{nl}\sim 37$ and $\Delta f^{(ort)}_{nl}\sim 25$.
More recently, by exploiting publicly available all--sky simulations,
\citet{lacasa2012} have tried to characterize the configuration
dependence and the frequency behavior of the bispectra originated by
``radio'' and ``far--infrared'' extragalactic sources. Their results
show that the bispectrum due to (extragalactic) far--infrared sources
starts to dominate that of radio sources on large angular scales at
150 GHz and dominates on the whole multipole range at 350 GHz.

In this paper we characterize the contamination that unresolved
extragalactic point sources introduce in the CMB bispectrum by using
the best available predictions on number counts of ``radio'' sources,
i.e. Active Galactic Nuclei (AGN) with emission spectra dominated by
synchrotron radiation (in this frequency range), and of
``far--infrared'' (far--IR) sources, i.e. low-- and high--redshift
galaxies in which the microwave emission is dominated by warm/cold
dust in thermal equilibrium with the radiation field. We also discuss
the effect of unresolved {\it polarized} extragalactic sources in
``future'' CMB anisotropy experiments. In particular, we make specific
predictions for different frequency channels foreseen for the future
Cosmic Origin Explorer (COrE) satellite mission \citep[]{core2011},
which has been recently submitted to the European Space Agency (ESA)
Horizon 2015-2025 Programme.\footnote{The COrE mission has not been
  selected by ESA in the 2011 first call for proposals. However, a
  COrE--like space mission with technical characteristics similar to
  the present ones could be hopefully funded in the near future, given
  that its very high scientific output has been already acknowledged
  by the ESA Space Science Advisory Committee (SSAC).  In fact, a
  COrE--like mission will be of fundamental scientific importance for
  detecting the imprint left by primordial gravitational waves on
  polarized CMB anisotropies, since the amplitude of primordial
  gravitational waves is directly proportional to the energy scale at
  which the inflation occurred. We don't know the exact instrumental
  characteristics of this future space mission. However, we do useful
  predictions by using the current instrumental technical details
  quoted in the proposal \citep{core2011}.}

Our current predictions of number counts of extragalactic sources are
based on two different models: the first one, by \citet{tucci2011}, is
able to give the most precise predictions on high--frequency ($\nu\geq
100$ GHz) number counts of radio sources using physically motivated
recipes to describe the complex spectral behavior of blazars, that
dominate the mm-wave counts at bright flux densities; the second one,
by \citet{lapi2011}, using a simple physical model and a single
reference Spectral Energy Distribution (SED), has proved able to
reproduce very well all the most recent measures (by Herschel, BLAST,
etc.) of number counts of extragalactic sources at sub-mm
wavelengths. It is important to mention that, to our knowledge, this
is the first time that predictions on the impact of extragalactic
radio and far-IR/sub-mm sources on the CMB angular power spectrum,
bispectrum and other statistics are performed with the level of detail
discussed here.

In Section \ref{sect:pointsources} we describe the extragalactic point
source population in this frequency range. The models that
characterise their power spectrum are presented in Section
\ref{sect:powerspectra}. Section \ref{sect:primnongauss} reviews the
basic formalism for the angular bispectrum and the four bispectrum
shapes.  Section \ref{sect:bias} introduces the point source bispectra
and the estimator to measure the $\Delta f_{nl}$ bias. The forecasts
for the COrE mission are given in Section \ref{sect:results} and,
finally, in Section \ref{sect:conclusions} the conclusions are listed.
\section{Extragalactic Point Sources}
\label{sect:pointsources}
In this section we describe the expected contribution of extragalactic
point sources to temperature and polarization fluctuations from
centimetre to sub-millimetre wavelengths. In this range of
frequencies, extragalactic sources are typically divided in two
populations, according to the physical radiative mechanism involved:
far--infrared (far--IR) sources dominated by thermal dust emission and
radio sources dominated by synchrotron emission. In the following
analysis we use number counts predictions obtained by two very recent
models: i.e., \citet[][T11, hereafter]{tucci2011}, as for radio
sources at cm to mm wavelengths and \citet[][L11, hereafter]{lapi2011},
as for far--IR/submm sources. In both cases, these models provide
predictions on source number counts and related statistics which have
shown a very good agreement with observational data.
\subsection{Radiosources}
Recent evolutionary models for Extragalactic Radio Sources (ERS) by
\citet{zotti2005} and \citet{massardi2010} 
are capable of providing a good fit to number counts
from $\sim100$\,MHz and up to $\sim100$\,GHz. They
adopt a simple power-law, i.e. $S(\nu)\propto \nu^\alpha$, with an
almost flat spectral index ($\alpha\simeq -0.1$) for extrapolating
spectra of blazar sources\footnote{Blazars are jet-dominated
  extragalactic objects, observed within a small angle of the jet axis
  and characterized by a highly variable non-thermal synchrotron
  emission at GHz frequencies, in which the beamed component dominates
  the observed emission \citep{angel1980}} at high frequencies. These
models, however, over-predict the number counts of extragalactic
synchrotron sources detected by the Atacama Cosmology Telescope (ACT)
at 148\,GHz \citep{marriage2011} and by Planck in all the High Frequency
Instrument (HFI) channels
\citep{planck_536_13,planck_pip_38}. 
The main reason for this
disagreement is the spectral steepening observed in Planck Early
Release Compact Source Catalogue (ERCSC) sources at above $\sim70$ GHz
\citep{planck_536_7,planck_536_15} and already suggested by other data
sets \citep{gonzaleznuevo2008,sadler2008}.

The more recent models by T11 use physically grounded recipes to
describe the complex spectral behaviour of blazars, which dominate the
mm-wavelengths number counts at bright flux densities. The main
novelty of these models is the statistical prediction of a break
frequency, $\nu_M$, in spectra of blazars in agreement with classical
physical models for the synchrotron emission produced in blazar jets
(e.g., \citealt{blandford79}). The most successful of these models is
referred to as C2Ex\footnote{It assumes different distributions of
  break frequencies for BL\,Lac objects and Flat Spectrum Radio
  Quasars (FSRQs) -- with the relevant synchrotron emission coming
  from more compact regions in jets of the former objects. See T11 for
  more details.}. This model is able to give a very good fit to all
the data of bright extragalactic radio sources available so far:
number counts up to $\sim$500-600\,GHz and on spectral index
distributions up to, at least, 200-300 GHz \citep[see T11
  and][]{planck_pip_38,lopez2012}.

Polarization in ERS is typically observed to be a few percent of total
intensity at cm or mm wavelengths (e.g.,
\citealt{murphy2010,battye2011,sajina2011}), with only very few
objects showing a fractional polarization, $\Pi=P/S$, as high as
$\sim10\%$. We refer to the recent paper by \citet{tucci2012} in order
to characterize polarization properties of ERS and their contribution
to power spectra. In that work authors show that the distribution of
the fractional polarization for blazars is well described by a
log--normal distribution with values of the median fractional
polarization, $\Pi_{med}$, and of $\langle\Pi^2\rangle^{1/2}$
depending on the frequency. 
At frequencies $\nu\ga40\,$GHz we take $\Pi_{med}=0.04,\,0.03,\,0.036$
and $\langle\Pi^2\rangle^{1/2}=0.06,\,0.038,\,0.045$ for
steep--spectrum sources, FSRQs and BL\,Lacs, respectively. 
These values are intermediate between the two "extreme" cases
discussed in \citet{tucci2012} and rely on the most recent data at
86\,GHz of \citet{agudo2010}.
\subsection{Far-IR Sources}
Recent data acquired by The Herschel Astrophysical TeraHertz Large
Area Survey \citep[H-ATLAS][]{eales2010}, the largest area survey
carried out by the Herschel Space Observatory \citep{pilbratt2010},
have demonstrated that far-IR sources comprise both a low--z galaxy
population, identified through matching to the Sloan Digital Sky
Survey \citep{york2000} data \citep{smith2011a}, and a high-z
population (median redshift $\sim$2) identified through their far-IR
colors \citep{amblard2010}. Low-z galaxies are generally
normal/star-forming late-type galaxies with moderate opacity
\citep{dunne2011,smith2012}. Through analyses of clustering these two
populations are found to be very different; the low-z population
($z<0.3$) does cluster like star--forming blue galaxies
\citep{maddox2010,guo2011,kampen2012}, whereas the high--z population
clusters much more strongly, suggesting that the high-z sources reside
in more massive halos \citep{maddox2010}.

In the case of late--type infrared galaxies, we adopt the same source
number counts as in L11, previously derived by
\citet{negrello2007}. As explained in more detail in that paper, the
contributions of the this type of galaxies to the number counts were
estimated following a phenomenological approach, which consists of
simple analytic recipes to evolve backwards their local LFs with
cosmic time, as well as of appropriate templates for their SEDs, to
compute K-corrections.

Regarding star-forming proto-spheroidal galaxies, the 100 and 250
$\mu$m luminosity functions (LF) at different redshifts are quite well
reproduced by the physical model of Early Type Galaxy (ETG) formation
and evolution by \citet{granato2001,granato2004}, further elaborated
by \citet{lapi2006}, and recently revised by L11 without any
adjustment of the parameters. As discussed in these papers, the model
is built in the framework of the standard hierarchical clustering
scenario.  This model was the basis for the successful predictions of
the submillimetre counts of strongly lensed galaxies by
\citet{negrello2007}. It also accurately reproduced the
epoch-dependent galaxy LFs in different spectral bands, as well as a
variety of relationships among photometric, dynamical, and chemical
properties, as shown in previous papers \citep[see Table 2
  of][]{lapi2006} and additional results, especially on the galaxy
chemical evolution, in \citet{mao2007,lapi2008}.

Number counts from the L11 model are observed to be slightly high at
mm wavelengths\footnote{This effect is due to the fact that the SED of
  high redshift objects (that contribute to the mm counts around
  10-30\,mJy) are slightly different to the template adopted in L11 as
  a universal SED for protospheroids.}. The model, in fact, tends to
overestimate the Poisson contribution to the power spectrum of
intensity fluctuations measured at frequencies $\le353\,$GHz by {\it
  Planck} \citep{planck_536_18}, SPT \citep{hall10} and ACT
\citep{dunkley11}. Scaling down counts by constant factors is however
enough to comply with measurements at mm wavelengths
\citep{xia2012}. In this work we multiply number counts by the
following factors: 0.81 and 0.71 at 375 and 225\,GHz, 0.55 for
$\nu\le165$\,GHz.
\begin{figure}
\includegraphics[width=84mm]{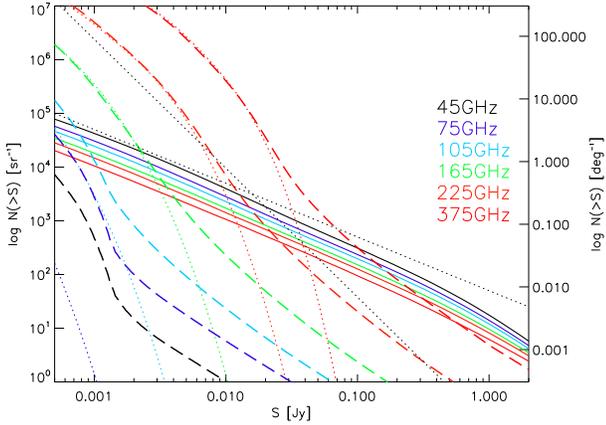}
\caption{Predicted integral number counts of radio sources
   (solid lines; T11 model) and of far-IR sources (total
    counts: dashed lines; proto-spheroidal galaxies only:
    dotted lines; L11 model) at the frequencies indicated
    inside the panel (each frequency and the corresponding
    curves are displayed by a different colour). The two
    black dotted lines, plotted as a reference, represent
    two power laws, with slopes 1 and 3, respectively.}
\label{f1}
\end{figure}

Finally, as predicted by
\citet{blain1996,perrotta2002,perrotta2003,negrello2007,paciga2009,lima2010}
among others, the combination of the very steep number counts in the
far--IR/sub-mm and of high redshifts -- hence large lensing optical
depths -- maximizes the fraction of strongly lensed sources. In the
case of the L11 model, the contribution from lensed proto--spheroidal
galaxies were estimated using the amplification distribution of
\citet{perrotta2003,negrello2007,lapi2012}. Such predictions were
recently confirmed by \citet{negrello2010,gnuevo2012} by using the
Science Demonstration Phase data of the H-ATLAS survey. For the
current work we have considered that the strong gravitational lensing
effect is a random process, and therefore, the lensed proto-spheroidal
galaxies show a Poisson spatial distribution. Due to their lack of
correlation with the un-lensed proto-spheroidal galaxy sky
distribution, we have treated them as a different population.

Regarding polarization, all the - very few - data so far collected on
the polarization of far-IR/sub-mm sources (dusty early- or late-type
galaxies and Galactic dust clouds) do indicate that the total
polarized emission of these sources has to be very low with median
values at around 1-2\% level. Moreover, in many observations, no
polarized emission (or compatible with zero) has been detected
\citep{dotson2010,vaillan2011}. Due to the above, we have assumed --
for the population of far-IR selected sources as a whole -- a
(conservative) average polarization level of 1\%.
\section{Contamination of point sources to CMB power spectra}
\label{sect:powerspectra}
In Fig. \ref{f1} we directly compare expected integrated number counts
of radio and far-IR sources from the T11 and L11 models at frequencies
ranging from 45 to 375\,GHz -- i.e. the frequencies interesting for
CMB studies. The figure well emphasizes the different behavior of the
two populations in terms of flux density distribution and frequency
spectrum. In the case of ERS, we can notice that the slope of integral
counts ($N(>S)\propto S^{-\beta}$) is not very different from
$\beta\simeq1$ at $S<1\,$Jy. This means that the contamination of
unresolved ERS to the CMB power spectra is almost proportional to the
flux density limit, $S_c$, above which sources can be detected and removed or
masked. Moreover, because the spectra of blazars are typically flat or
moderately steep, $-1\la\alpha\la0$, integral counts change slowly
with the frequency, decreasing by only a factor 2--4 from 44 to
350\,GHz, depending on the flux density considered.

\begin{figure*}
\includegraphics[width=160mm]{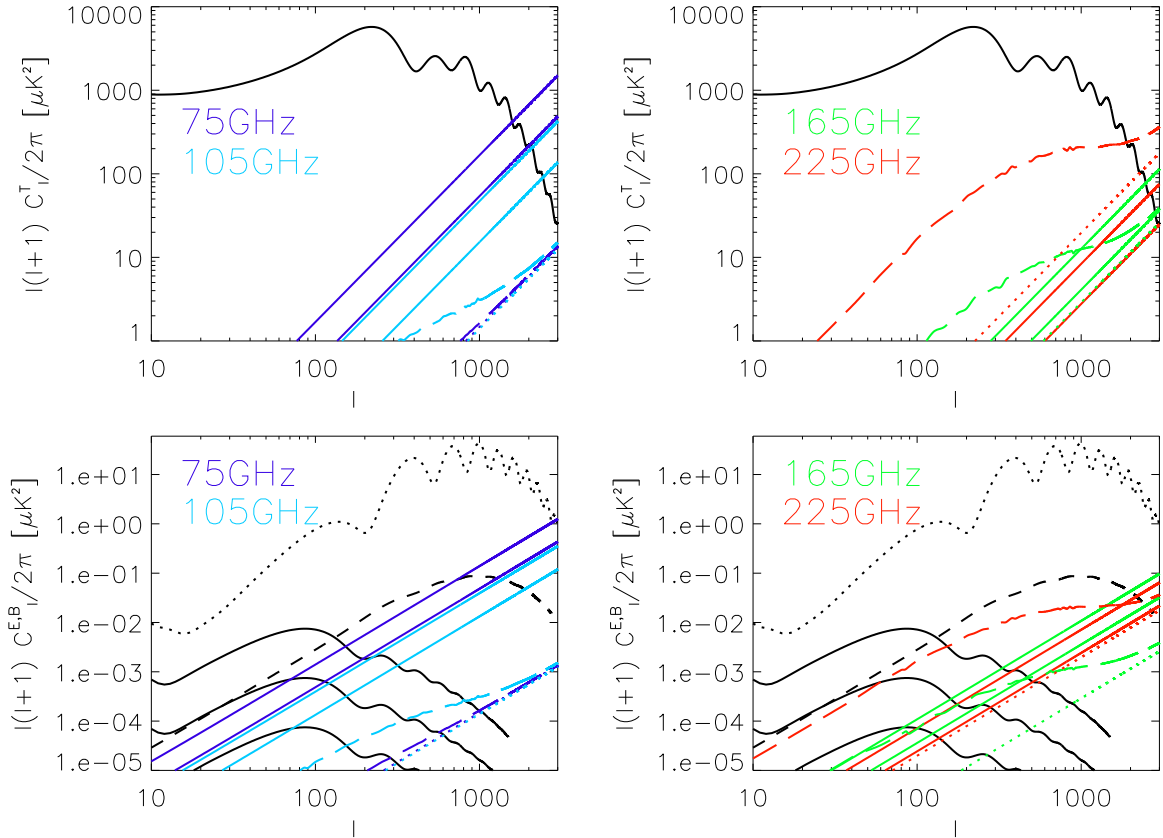}
\caption{Point source power spectra for temperature (top panels) and
  polarization (bottom panels) at 75, 105, 165 and 225 GHz. Lines in
  colour are for radio sources (solid lines) and for far-IR sources
  (dashed lines for the total contribution; dotted lines for the
  Poisson contribution). We consider two flux density limits for
  radio sources: $S_c=$1 and 0.3 Jy; for far-IR sources we plot only the
  case with $S_c=$1 (there are not significant differences taking
  $S_c=0.3$). Black lines are for CMB spectra from the fiducial WMAP
  7-year model: in the {\it bottom panels} dotted lines are for the
  E--mode, dashed lines for lensing--induced B--mode and solid lines
  for primordial B--mode with $r=0.1,\,0.01,\,10^{-3}$.}
\label{f2}
\end{figure*}

On the other hand, far-IR source counts 
increase very steeply when moving to low flux densities, due to the contribution of
proto--spheroidal galaxies 
(see dotted lines in
Fig.\,\ref{f1}). The general slope of far-IR integral counts, $\beta$, is
between 2 and 3 depending on the flux density range. As a consequence,
the amplitude of power spectra for unresolved far-IR sources is almost
insensitive to the value of $S_c$, at least for flux density limits typically
obtained in large--area CMB experiments. Finally, because of the steep
frequency spectrum of dust emission, $N(>S)$ is also strongly
dependent on frequency, making the contribution of far-IR sources already
completely negligible at frequencies $\nu\la100\,$GHz (see Fig.
\ref{f2}).

In terms of power spectra, it is well known that point sources without spatial
correlations have:
\begin{equation}
C_{\ell}^{white} = k_{\nu}^2 \int_{0}^{S_c}S^2 \frac{dn}{dS}dS
\label{white_s_cls}
\end{equation}
where $k^{-1}_{\nu}=dB_0/dT_0$ is the conversion factor from flux density
to temperature units \citep[see e.g.][]{tegmark96}, $S_c$ is the flux density
limit for the unresolved point sources and $dn/dS$ is number counts
that characterise the population of sources. However, powerful high--z
far-IR sources have been shown to be highly correlated by, e.g.,
recent measures from 
{\it Herschel} \citep{amblard11,viero12} 
and {\it Planck} \citep{planck_536_18}. For the other extragalactic
source populations the clustering is negligble 
\citep{toffolatti2005,negrello2007}. Therefore, we shall consider the
contribution of clustered sources to angular power spectra (and
bispectra) only for dusty proto--spheroidal galaxies. This is
estimated from realistic simulations of two--dimensional distributions
of clustered point sources \citep[see][]{gonzaleznuevo2005}. 
This software was also implemented in the Planck Sky Model
\citep{delabrouille2012} and its predictions on the power spectrum
of cosmic infrared background anisotropies have been shown to be in
good agreement with the latest Planck measurements
\citep{planck_536_18}.

The total power spectra from point sources are then given by the sum
of the following components:
\begin{equation}
C_{\ell}^{ps} = C_{\ell}^{rad,white}+C_{\ell}^{ir,white}+C_{\ell}^{sph,clust}
\label{eq:total_ps_cl_bispectrum}
\end{equation}
where $C_{\ell}^{rad,white}$ and $C_{\ell}^{ir,white}$ are the
constant term from shot-noise power spectra of radio and far-IR (which
includes late--type, lensed and proto--spheroidal galaxies) sources,
and $C_{\ell}^{sph,clust}$ is the power spectrum from the clustered
contribution of proto--spheroidal galaxies\footnote{We remind the
  reader that the total power spectrum of a population of clustered
  sources is $C_{\ell}^{tot}=C_{\ell}^{white}+C_{\ell}^{clust}$.}.

Regarding polarization, point sources contribute, on average, equally
to power spectra of E-- and B--modes or equivalently to $Q$ and $U$
power spectra in the flat--sky limit.  Polarization power spectra of
point sources are proportional to the temperature power spectrum
\citep{tucci04}: 
for the $Q$
power spectrum, e.g., we have
\begin{equation}
C _{\ell}^{Q} = k_{\nu}^2\,\langle S^2\,p^2\,\cos^2(2\phi) \rangle
= {k_{\nu}^2 \over 2}\,\langle p^2 \rangle\,C^{T}_{\ell}\,.
\end{equation}
The above relation assumes that the flux density $S$, the polarization
fraction $p$ and the polarization angle in the chosen reference system
$\phi$ are independent variables \citep[see][for a detailed discussion
on this assumption]{tucci2012}. On the other hand, all the cross-power
spectra are equal to zero:
\begin{center}
\begin{align}
C _{\ell}^{TQ} & \propto \langle S^2\,p\,\cos(2\phi) \rangle=0
\nonumber \\
  C _{\ell}^{TU} & \propto \langle S^2\,p\,\sin(2\phi) \rangle=0
\nonumber \\
 C _{\ell}^{QU} & \propto \langle S^2\,p^2\,\cos(2\phi)\,\sin(2\phi)
\rangle=0\,.
\end{align}
\end{center}
Fig.\,\ref{f2} shows model predictions for temperature and
polarization power spectra at 75, 105, 165 and 225\,GHz.  We also
consider two possible flux density limits, i.e. $S_c=1$ and 0.3\,Jy.

As previously discussed, far-IR spectra are almost insensitive to the
choice of $S_c$, especially for the clustered contribution. Comparing
the two source populations, we see that radio sources are dominant at
$\nu\la100\,$GHz, while far-IR sources start to dominate at
$\nu\ga200$\,GHz both in temperature and polarization.  At
intermediate frequencies, i.e. $100<\nu<200$\,GHz, ERS are the main
contaminant at high multipoles ($\ell>1000$), but not for
$\ell\la1000$, where the far-IR clustered sources contribution becomes
higher than the ERS power spectrum in temperature and comparable in
polarization.

\begin{figure}
\includegraphics[width=84mm]{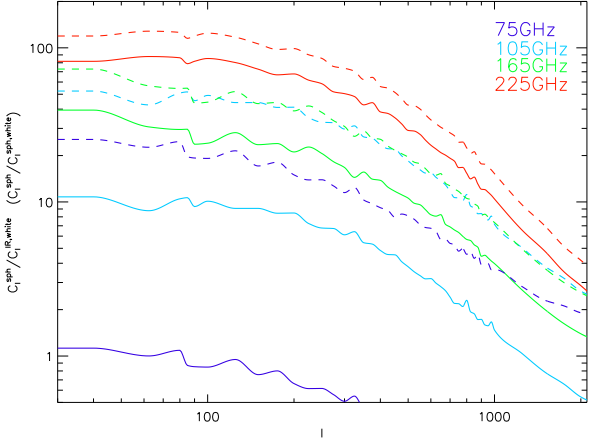}
\caption{The ratio of the total spectrum, $C_{\ell}^{sph}$, to the
  shot--noise spectrum, $C_{\ell}^{sph,white}$, of proto--spheroidal
  galaxies (dashed lines), and to the shot--noise spectrum from far-IR
  sources, $C_{\ell}^{ir,white}$ (solid lines), at the frequencies
  indicated in the panel.}
\label{f3}
\end{figure}

Because proto--spheroidal galaxies are the only clustered population
in the L11 model, it is worth discussing more in detail their
relevance in terms of power spectrum. In Fig.\,\ref{f3} we plot the
ratio of the total spectrum of proto--spheroidal galaxies to their
shot--noise spectrum, $C_{\ell}^{sph}/C_{\ell}^{sph,white}$, and to
the shot--noise spectrum of all far-IR sources,
$C_{\ell}^{sph}/C_{\ell}^{ir,white}$, (the former ratio enters later
in Eq.\,\ref{eq:total_sph_bipectrum} and gives us the increase of the
bispectrum due to clustering). As the frequency decreases, we observe
two effects: {\it a}) the contribution of proto--spheroidal galaxies
to the far-IR power spectrum becomes less and less important; {\it b})
moreover, the relative contribution of clustering to proto--spheroidal
power spectrum reduces. The reason of these trends is related to the
shift of the redshift distribution of submillimetre bright galaxies to
higher redshift with increasing the wavelength, as effect of the
strongly negative $K$--correction (L11). At high z, however, massive
proto--spheroidal galaxies become increasingly rare and the slope of
number counts at low frequencies is not as steep as that at sub-mm
wavelengths. As a consequence, the shot--noise term of the power
spectrum, dominated by bright sources, has more weight with respect to
the clustered one (dominated by faint sources).

Finally, we discuss the contamination of extragalactic sources to CMB
fluctuations: in the frequency range considered, it is relevant only
at very high $\ell$ ($\ga2000$) for temperature and is essentially
negligible for the E--mode polarization (see Fig.\,\ref{f2}). This is
not the case for the B--mode polarization arisen from primordial
gravitational waves if $r<0.01$. A subtraction of points sources at
levels even lower than ones shown in Fig.\,\ref{f2} is not an easy
task. For radio sources, this could be in principle done {\it a}) by
removing/masking sources down to flux density limits of $S<0.3$\,Jy or {\it
  b}) by a direct measurement of the power spectrum of polarization
fluctuations at small angular scales. In both cases observations with
arcmin resolution are required. Far-IR sources are still more
challenging to deal with, and the previous methods are not valid to
clean their contamination due to the high correlated signal. Instead,
approaches similar to ones used for diffuse Galactic foregrounds
should be more suitable for this goal, although they are model
dependent. 
%
%
\section{The angular bispectrum}
\label{sect:primnongauss}
The CMB angular bispectrum is defined in terms of cubic combinations
of the spherical harmonic coefficients of the CMB anisotropies of the
T, E and B modes:
\begin{equation}
B_{\ell_1 \ell_2 \ell_3, m_1 m_2 m_3}^{ijk}
\equiv \langle a^i_{\ell_1 m_1} a^j_{\ell_2 m_2}a^k_{\ell_3 m_3}\rangle,
\label{ang_bispI}
\end{equation}
where the indices $\{i,j,k\}$ cover the T, E and B modes. The
angular-averaged bispectrum and the reduced angular-averaged
bispectrum, $B^{ijk}_{\ell_1 \ell_2 \ell_3}$ and $b^{ijk}_{\ell_1 \ell_2 \ell_3}$
are defined as:
\begin{eqnarray}
\nonumber
& B^{ijk}_{\ell_1 \ell_2 \ell_3} = &\\
\nonumber
& \sum_{m_1 m_2 m_3} \left
  ( \begin{array}{ccc} \ell_1 & \ell_2 & \ell_3 \\
m_1 & m_2 &
  m_3 \end{array}\right ) B^{ijk}_{\ell_1 \ell_2 \ell_3, m_1 m_2 m_3} =\\
& b^{ijk}_{\ell_1\ell_2\ell_3}
\left ( \begin{array}{ccc} \ell_1 & \ell_2 &
  \ell_3 \\ 0 & 0 & 0 \end{array}\right )
\sqrt{\frac{(2\ell_1+1)(2\ell_2+1)(2\ell_3+1)}{4\pi}}. &
\end{eqnarray}
The reduced bispectrum can be written as \citep[see,
  e.g.][]{yadav2010}:
%
\bigskip\bigskip

\begin{align}
\nonumber
& b^{ijk}_{\ell_1 \ell_2 \ell_3} = (4\pi)^3(-i)^{\ell_1+\ell_2+\ell_3}\sum_{m_1 m_2 m_3} \left
  ( \begin{array}{ccc} \ell_1 & \ell_2 & \ell_3 \\ m_1 & m_2 &
  m_3 \end{array}\right ) \\
\nonumber
& \times \int \frac{d^3 k_1}{(2\pi)^3}\frac{d^3 k_2}{(2\pi)^3} \frac{d^3 k_3}{(2\pi)^3}
Y_{\ell_1 m_1}^*(\hat{k_1})Y_{\ell_2 m_2}^*(\hat{k_2})Y_{\ell_3 m_3}^*(\hat{k_3})\\
& g^i_{\ell_1}(k_1)g^j_{\ell_2}(k_2)g^k_{\ell_3}(k_3)\langle \Phi({\bf k_1}) \Phi({\bf k_2}) \Phi({\bf k_3}) \rangle,
\label{cmb_bispectrum}
\end{align}
where $g^i_{\ell}(k)$ is the transfer function for temperature or
polarization\footnote{This function can be computed numerically using
  the CAMB \citep{lewis2000} software http://camb.info} and $\langle
\Phi({\bf k_1}) \Phi({\bf k_2}) \Phi({\bf k_3}) \rangle$ is the
3-point correlation function in Fourier space of the Bardeen's
curvature perturbations $\Phi({\bf x})$. The 3-point correlation
function can be given in terms of the so-called shape function
$F(k_1,k_2,k_3)$:
\begin{equation}
\langle
\Phi({\bf k_1}) \Phi({\bf k_2}) \Phi({\bf k_3}) \rangle =
(2\pi)^3\delta^3({\bf k_1+k_2+k_3})F(k_1,k_2,k_3)
\label{shape_function}
\end{equation}
The level of primordial non-Gaussianity is parametrised by the
non-linear coupling parameter $f_{nl}$
\citep{verde2000,komatsu2001,bartolo2004,yadav2010}. This parameter
measures departures from zero of the bispectrum. Depending on the
physical mechanisms of the different inflationary models the shape
function can take different forms. The shapes that we consider in this
work are the local, equilateral, orthogonal and flat shapes.
\begin{itemize}
\item {\bf Local shape.} This type of Non-Gaussianity is generated in
  multi-field inflationary models \citep{komatsu2005, komatsu2010b},
  the curvaton \citep{lyth2003}, the inhomogeneous reheating scenario
  \citep{dvali2004,bartolo2004}, hybrid inflation \citep{lin2009},
  etc. Its shape is given by \citep[see
    e.g.][]{creminelli2006,yadav2010}:
\begin{equation}
F(k_1,k_2,k_3)=2A^2f_{nl}\Big[\frac{1}{k_{1}^{3-(n_s-1)}k_{2}^{3-(n_s-1)}}+(2~perm)\Big],
\label{local_shape}
\end{equation}
where $A$ is the amplitude of the power spectrum and $n_s$ is the
spectral index \citep{komatsu2011}. Most of the signal of this type of
bispectrum is located in squeezed configurations such as $k_1 \simeq
k_2 \gg k_3$ \citep{creminelli2006,yadav2010,martinez2012}.
\item {\bf Equilateral shape.}  This type of Non-Gaussianity is
  generated in the Dirac-Born-Infeld inflation
  \citep{silverstein2004,bartolo2004,langlois2008}, ghost inflation
  \citep{arkani-hamed2004}, single-field inflationary models in
  Einstein gravity \citep{chen2007} etc. Its shape is given by
  \citep[see e.g.][]{creminelli2006,yadav2010}:
\begin{align}
\nonumber
F(k_1,k_2,k_3)& =6A^2f_{nl}\Big[-\frac{1}{k_{1}^{3-(n_s-1)}k_{2}^{3-(n_s-1)}} +(2~perm)\\
\nonumber
& -\frac{2}{(k_1k_2k_3)^{2(4-n_s)/3}} \\
& + \Big\{\frac{1}{k_1^{(4-n_s)/3}k_2^{2(4-n_s)/3}k_3^{(4-n_s)}} + (5~perm)\Big\}\Big].
\label{equilat_shape}
\end{align}
Most of the signal of this type of bispectrum is located in
equilateral configurations such as $k_1 \simeq k_2 \simeq k_3$
\citep[see e.g.][]{creminelli2006,yadav2010,martinez2012}.
\item {\bf Orthogonal shape.} This type of Non-Gaussianity is
  generated in general single-field models
  \citep{cheung2008,senatore2010}. Its shape is given by \citep[see
    e.g.][]{senatore2010,yadav2010,komatsu2011}:
\begin{align}
\nonumber
F(k_1,k_2,k_3)&=6A^2f_{nl}\Big[-\frac{3}{k_{1}^{3-(n_s-1)}k_{2}^{3-(n_s-1)}}+(2~perm)\\
\nonumber
&-\frac{8}{(k_1k_2k_3)^{2(4-n_s)/3}} \\
& +\Big\{\frac{3}{k_1^{(4-n_s)/3}k_2^{2(4-n_s)/3}k_3^{(4-n_s)}} + (5~perm)\Big\}\Big].
\label{ortho_shape}
\end{align}
This bispectrum peaks in both equilateral $k_1 \simeq k_2 \simeq k_3$
and flat-triangle configurations such as $k_1 = k_2 + k_3$
\citep[see the shape figures
  e.g. in][]{senatore2010,martinez2012}.
\item {\bf Flat shape.} This type of Non-Gaussianity is generated for
  example in models with modifications in the initial state of the
  inflaton field \citep[see e.g.][and references
    therein]{meerburg2009}, single-field models with high derivative
  iterations \citep{bartolo2010b} among others. Its shape is given by
  \citep{meerburg2009}:
\begin{align}
\nonumber
F(k_1,k_2,k_3)& =6A^2f_{nl}\Big[\frac{1}{k_{1}^{3-(n_s-1)}k_{2}^{3-(n_s-1)}}+(2~perm)\\
\nonumber
& +\frac{3}{(k_1k_2k_3)^{2(4-n_s)}} \\
& -\Big\{\frac{1}{k_1^{(4-n_s)/3}k_2^{2(4-n_s)/3}k_3^{(4-n_s)}} + (5~perm)\Big\}\Big].
\label{flat_shape}
\end{align}
This bispectrum peaks in flat-triangle configurations $k_1 = k_2 +
k_3$ \citep[see the shape figure in][]{meerburg2009}.
\end{itemize}
Using the shape forms defined in
Eqs. \ref{local_shape}, \ref{equilat_shape}, \ref{ortho_shape} ,
\ref{flat_shape} and the CMB bispectrum in Eq. \ref{cmb_bispectrum} we
have:
\begin{align}
\big( b^{loc}_{\ell_1 \ell_2 \ell_3}\big)^{ijk} = 2\int_{0}^{\infty}x^2dx\Big[\alpha^i_{\ell_1}(x)\beta^j_{\ell_2}(x)\beta^k_{\ell_3}(x)  + (2~perm)\Big],
\label{localbispectrum}
\end{align}
\begin{align}
\nonumber
\big( b^{eq}_{\ell_1 \ell_2 \ell_3}\big)^{ijk} = 6 \int_{0}^{\infty}dxx^2\Big [ -\alpha^i_{\ell_1}(x)\beta^j_{\ell_2}(x)\beta^k_{\ell_3}(x) + (2~perm)\\
 + \beta^i_{\ell_1}(x)\gamma^j_{\ell_2}(x)\delta^k_{\ell_3}(x)  + (5~perm) -2 \delta^i_{\ell_1}(x)\delta^j_{\ell_2}(x)\delta^k_{\ell_3}(x) \Big ],
\label{equilateralbispectrum}
\end{align}
\begin{align}
\nonumber
\big( b^{ort}_{\ell_1 \ell_2 \ell_3}\big)^{ijk} = 18 \int_{0}^{\infty}dxx^2\Big [ -\alpha^i_{\ell_1}(x)\beta^j_{\ell_2}(x)\beta^k_{\ell_3}(x) + (2~perm) \\
+ \beta^i_{\ell_1}(x)\gamma^j_{\ell_2}(x)\delta^k_{\ell_3}(x) + (5~perm) -\frac{8}{3} \delta^i_{\ell_1}(x)\delta^j_{\ell_2}(x)\delta^k_{\ell_3}(x) \Big ]
\label{orthogonalbispectrum}
\end{align}
and
\begin{align}
\nonumber
\big( b^{flat}_{\ell_1 \ell_2 \ell_3}\big)^{ijk} = 6  \int_{0}^{\infty}dxx^2\Big [  \alpha^i_{\ell_1}(x)\beta^j_{\ell_2}(x)\beta^k_{\ell_3}(x) + (2~perm) \\
- \beta^i_{\ell_1}(x)\gamma^j_{\ell_2}(x)\delta^k_{\ell_3}(x) + (5~perm) + 3 \delta^i_{\ell_1}(x)\delta^j_{\ell_2}(x)\delta^k_{\ell_3}(x) \Big ]
\label{flatbispectrum}
\end{align}
where $\alpha^i_{\ell}(x)$, $\beta^i_{\ell}(x)$, $\gamma^i_{\ell}(x)$,
$\delta^i_{\ell}(x)$ are filter functions defined in terms of the
power spectrum $P(k)$ and the transfer functions $g^i_{\ell}(k)$
\citep[see e.g.][for their actual
  definitions]{fergusson2010a,fergusson2010b,yadav2010,curto2011b}.
\section{Bias and uncertainties on primordial non-Gaussianity}
\label{sect:bias}
\subsection{Bispectrum of point sources}
\begin{figure*}
\centering
\includegraphics[width=7.1cm, height=5.cm, angle= 0]{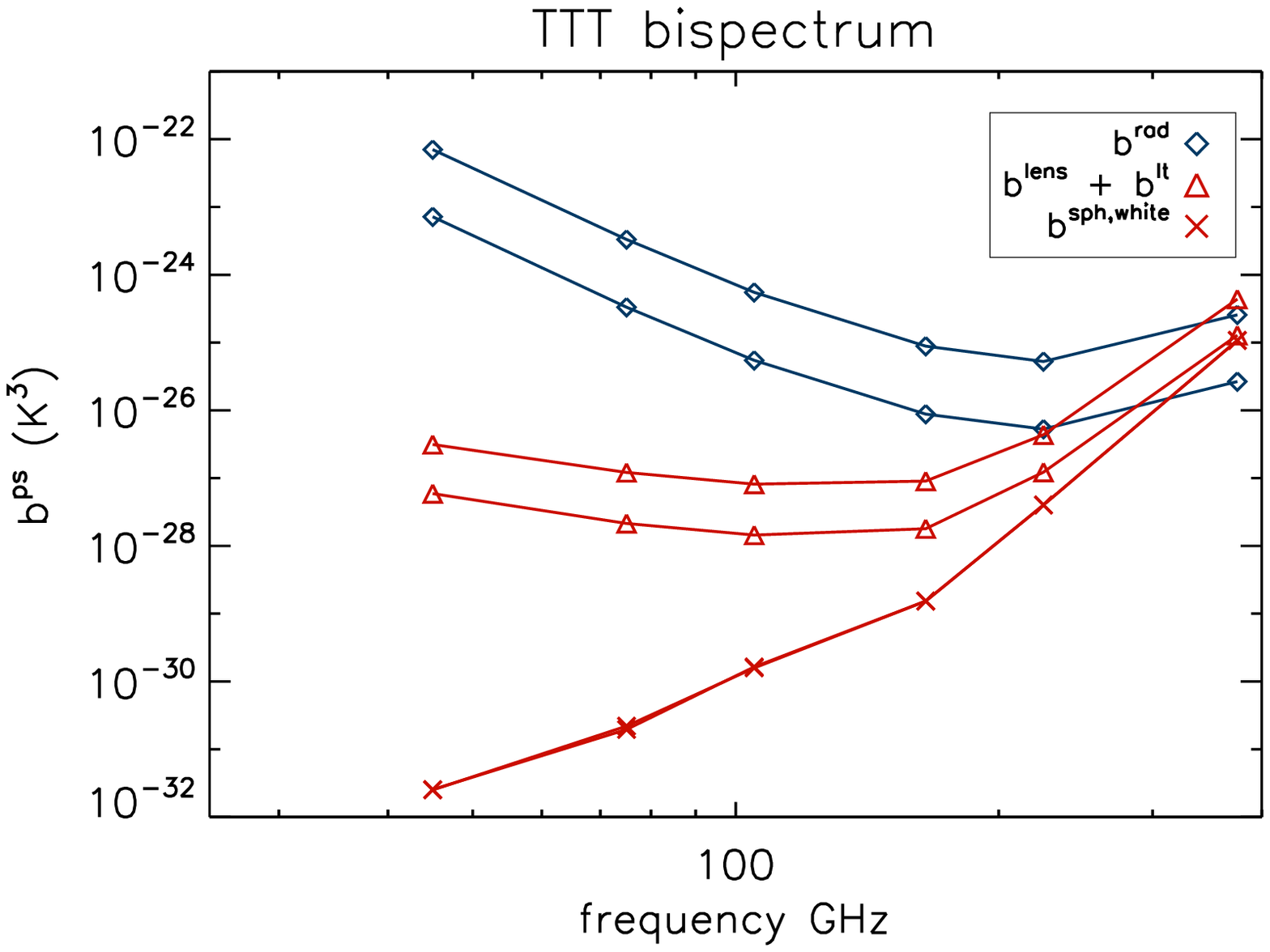}
\includegraphics[width=7.1cm, height=5.cm, angle= 0]{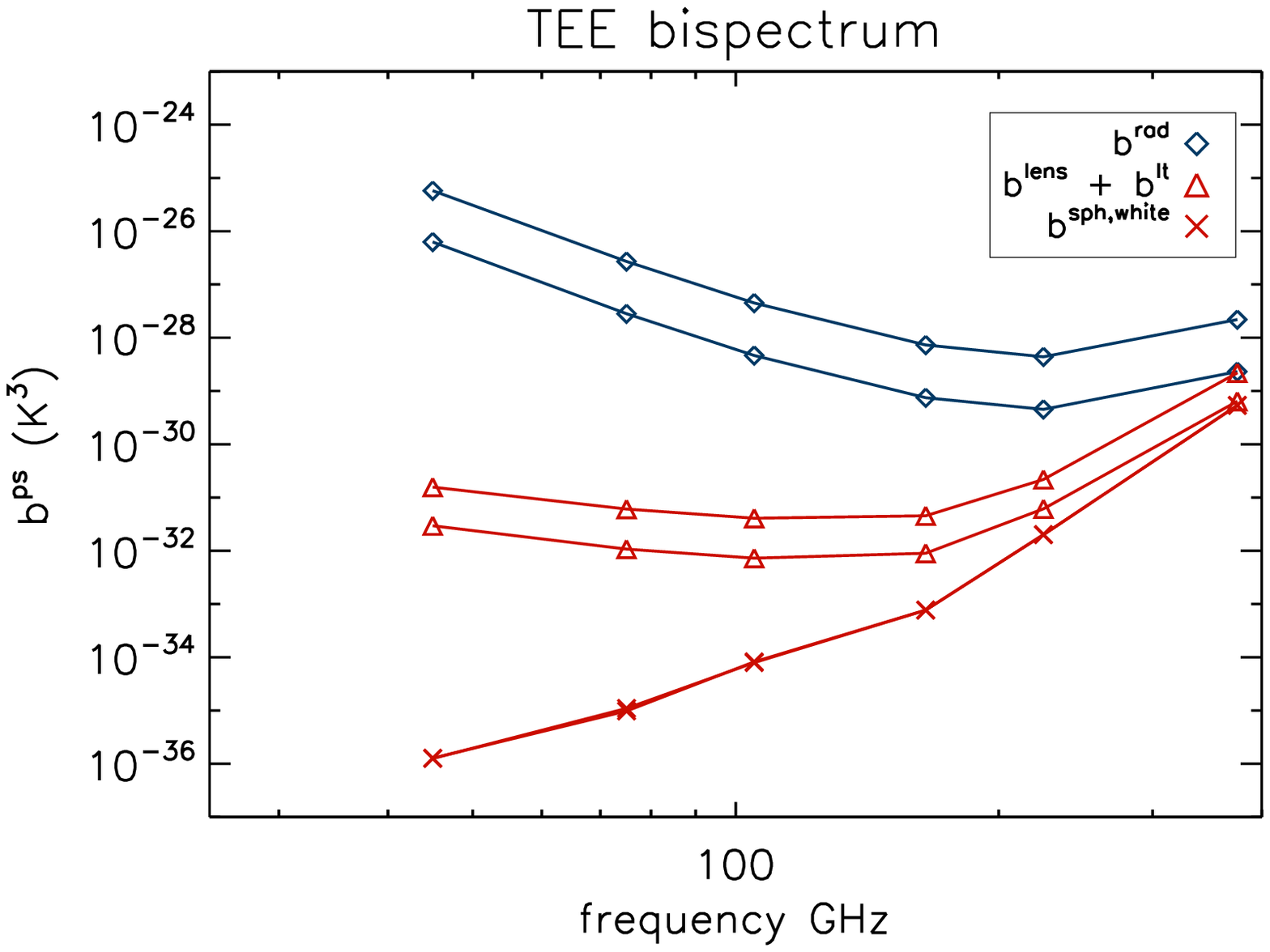}
\caption{\label{bispectrum_point_sources} The amplitude of the
  shot--noise bispectrum of radio sources ($b^{rad}$) and of the far-IR
  source (for late-type plus lensed galaxies, $b^{lt+le}$, and for
  proto-spheroidal galaxies, $b^{sph,white}$) as a function of
  frequency and for the two considered flux density limits, i.e. $S_c=0.3$
  and 1\,Jy. The left panel corresponds to the TTT bispectrum and the
  right panel corresponds to the TEE bispectrum.}
\includegraphics[width=7.1cm, height=5.cm, angle= 0]{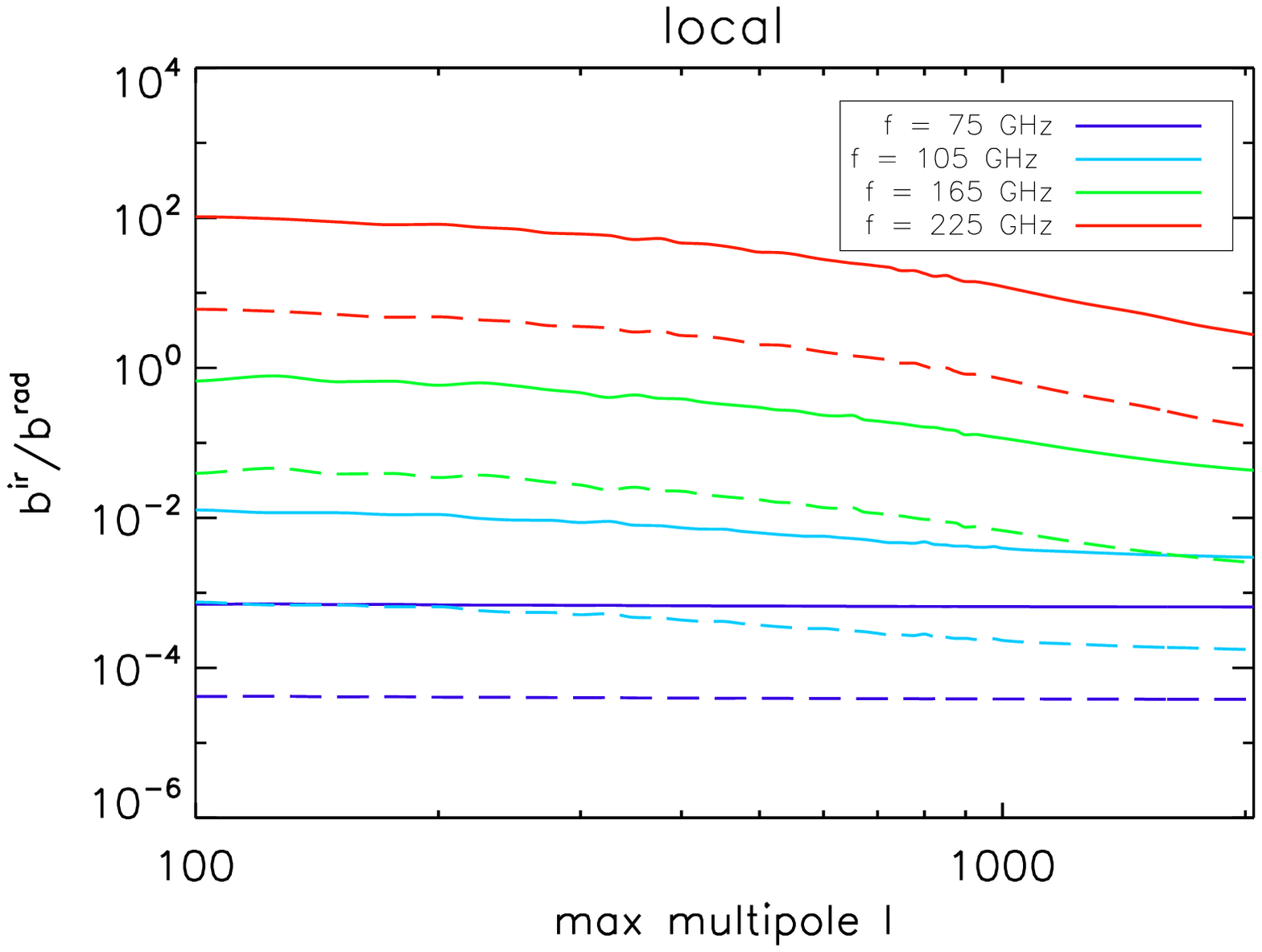}
\includegraphics[width=7.1cm, height=5.cm, angle= 0]{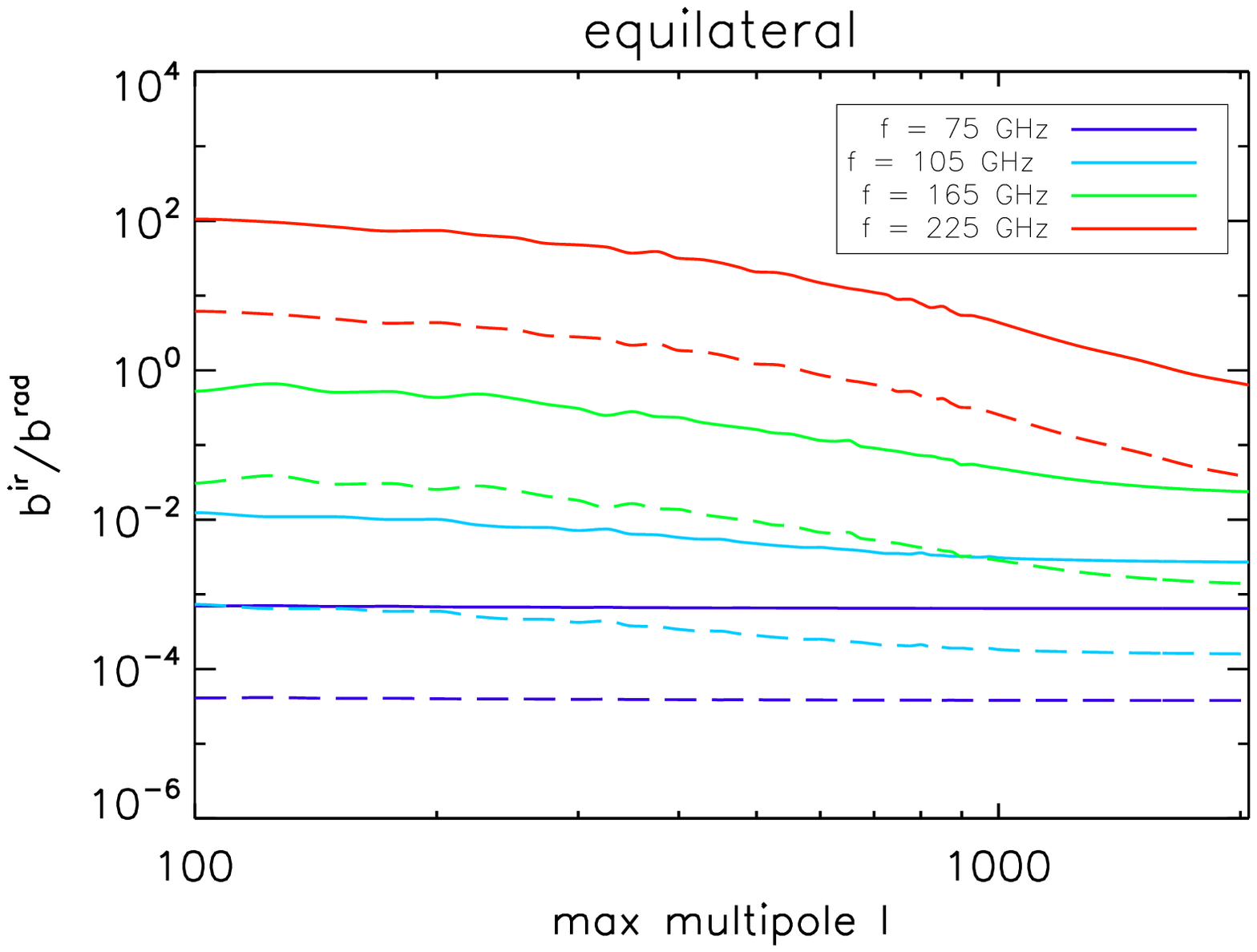}
\includegraphics[width=7.1cm, height=5.cm, angle= 0]{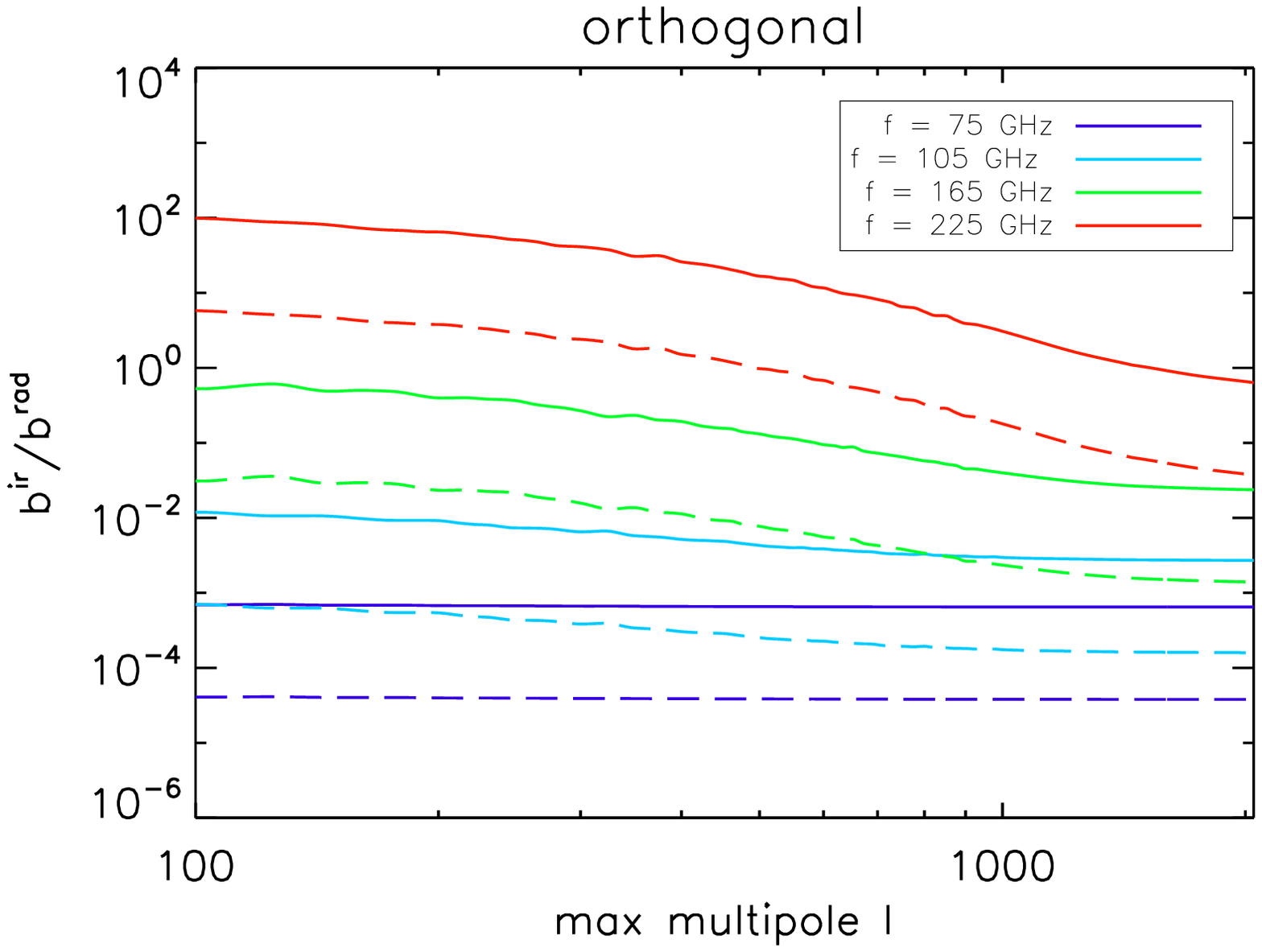}
\includegraphics[width=7.1cm, height=5.cm, angle= 0]{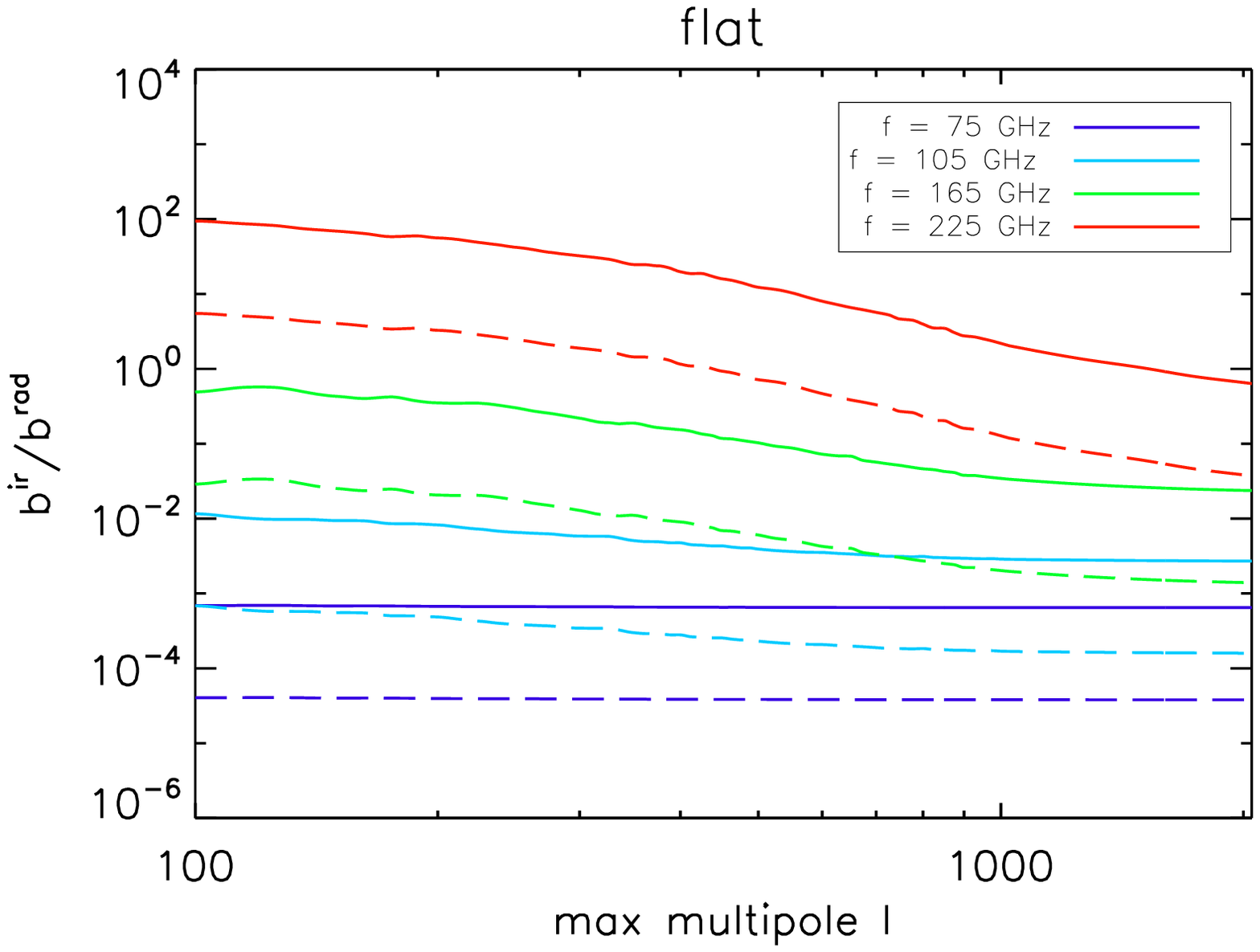}
\caption{\label{ratio_bispectrum_point_sources} Ratio of the far-IR
  source bispectrum $b^{ir}$ to the radio source bispectrum $b^{rad}$
  for $S_c~ =$ 0.3 Jy and the 4 considered shapes: local configuration
  ($\ell_1=\ell_2=\ell$ and $\ell_3=\ell_{min}=2$), equilateral
  configuration ($\ell_1=\ell_2=\ell_3=\ell$), isosceles orthogonal
  configuration ($\ell_1=\ell_2=\ell$ and $\ell_3= \sqrt{2} \ell)$ and
  isosceles flat configuration ($\ell_1=\ell_2=\ell$ and $\ell_3= 2
  \ell)$. The solid lines correspond to T, and the long dashed lines
  correspond to T and E.}
\end{figure*}
The bispectrum of a point source population without spatial
correlations, as the case of radio sources, is well characterised and
is constant with $\ell$:
\begin{equation}
b_{\ell_1 \ell_2 \ell_3}^{white} =
k_{\nu}^3\int_{0}^{S_c}S^3 \frac{dn}{dS}dS.
\label{white_s_b}
\end{equation}
On the other hand, far-IR sources have spatial correlations. According to
the widely used prescription by \citet{argueso2003}, also considered
e.g. in \citet{gonzaleznuevo2005,argueso2006} and extended to the full
sky by \citet{lacasa2012}, the total bispectrum for a single clustered
sources population can be written as:
\begin{equation}
b^{tot}_{\ell_1 \ell_2 \ell_3} \equiv
b _{\ell_1 \ell_2 \ell_3}^{white}\sqrt{\frac{
  C^{tot}_{\ell_1}C^{tot}_{\ell_2}C^{tot}_{\ell_3}}{
  C_{\ell_1}^{white}C_{\ell_2}^{white}C_{\ell_3}^{white}}}\,.
\label{eq:total_sph_bipectrum}
\end{equation}
The total bispectrum of point sources is then given by the sum of the
shot--noise contributions of ERS ($b_{\ell_1\ell_2\ell_3}^{rad,white}$)
and of far-IR late-type and lensed galaxies
($b_{\ell_1\ell_2\ell_3}^{lt+le,white}$) plus the ``clustered'' bispectrum of
proto--spheroidal galaxies ($b_{\ell_1\ell_2\ell_3}^{sph}$)
as given by Eq.\,\ref{eq:total_sph_bipectrum}:
\begin{align}
b^{ps}_{\ell_1\ell_2\ell_3} = & b_{\ell_1\ell_2\ell_3}^{rad,white} +
b_{\ell_1\ell_2\ell_3}^{lt+le,white} + b_{\ell_1\ell_2\ell_3}^{sph}\,.
\label{eq:total_ps_bispectrum}
\end{align}
%
When polarization is considered, we have 10 different bispectra arisen
from all the possible combinations of temperature and polarization
$E$-- and $B$--modes. However, on the base of the discussion in
Section\,\ref{sect:powerspectra}, it is easy to find out that the
cross--bispectrum $b^{TEB}_{\ell_1\ell_2\ell_3}$ and all bispectra
with an odd number of polarization terms are zero, i.e.
$b^{EEE}_{\ell_1\ell_2\ell_3}=b^{TTE}_{\ell_1\ell_2\ell_3}=...=0$. For the
remaining two cases, the bispectrum is proportional to
$b^{TTT}_{\ell_1\ell_2\ell_3}$ and it is given as:
\begin{equation}
b^{TEE}_{\ell_1 \ell_2 \ell_3} = b^{TBB}_{\ell_1 \ell_2 \ell_3} =
{1 \over 2}\langle p^2\rangle b^{TTT}_{\ell_1 \ell_2 \ell_3}.
\end{equation}
The total power spectrum and the bispectrum of the point sources are
the two ingredients needed to provide an estimate of their impact on
the primordial non-Gaussianity $f_{nl}$ parameter. The shot--noise
terms of $TTT$ and $TEE$ bispectra from radio and far-IR sources are shown
in Fig.\,\ref{bispectrum_point_sources}. The frequency at which far-IR
sources equal or overcome the ERS bispectrum is shifted to higher
frequencies with respect to the power spectrum, i.e. it is around
$300$ (350)\,GHz for the $TTT$ ($TEE$) bispectrum. This is in
agreement with the fact that sources with flux density close to $S_c$ give a
larger contribution to the bispectrum than to the power spectrum. A
significant dependence on the flux density limit is observed in the bispectrum
of ERS and of late--type plus lensed galaxies, but not of
proto--spheroidal galaxies.

Finally, we compare the total bispectrum of far-IR sources with the
ERS one as a function of multipoles and for the four used
configurations (see Fig.\,\ref{ratio_bispectrum_point_sources}). The
ratio is $\ga1$ only at 225\,GHz for $TTT$, and for $TEE$, but at low
multipoles. Otherwise, far-IR sources are always subdominant, or even
negligible, unless considering multipoles $\ell<100$. As widely
discussed in Sec.\,\ref{sect:powerspectra}, clustering of far-IR
sources becomes less relevant at low frequencies, and at
$\nu\la100$\,GHz the far-IR bispectrum is almost constant. Our
Fig.\,\ref{ratio_bispectrum_point_sources} can be directly compared to
Fig.\,9 of \citet{lacasa2012}. Various differences can be highlighted
from the two plots: 1) at high multipoles, the bispectrum ratio we
find is typically higher than one order of magnitude; 2) far-IR
bispectra by \citet{lacasa2012} have a steeper slope; 3) no frequency
dependence is observed in \citet{lacasa2012} results. These
discrepancies should be attributed to the different way to model
far-IR sources and their correlation.
\subsection{Bias and uncertainties on $f_{nl}$}
Considering weak levels of non-Gaussianity, the bias induced in the
primordial bispectrum $B^{prim}_{\ell_1 \ell_2 \ell_3}$ by the
point-source bispectrum $B^{ps}_{\ell_1 \ell_2 \ell_3}$ is
\citep[see e.g.][]{komatsu2001,bartolo2004,yadav2010,lacasa2012,martinez2012}:
\begin{equation}
\Delta f_{nl}  = \sigma^2\big(f_{nl}\big) \times \sum_{\ell_1 \le \ell_2 \le \ell_3 \le \ell_{max}}
\frac{B^{ps}_{\ell_1 \ell_2 \ell_3} B^{prim}_{\ell_1 \ell_2 \ell_3} }{\big(\Delta_{\ell_1 \ell_2
  \ell_3} C^{tot}_{\ell_1} C^{tot}_{\ell_2} C^{tot}_{\ell_3}\big)},
\label{eq:fnl_bias_point_sources}
\end{equation}
where
\begin{equation}
C^{tot}_{\ell} = \big( C^{CMB}_{\ell}+C^{ps}_{\ell}\big) w^2_{\ell} + C^{noise}_{\ell},
\label{total_cls}
\end{equation}
$C^{CMB}_{\ell}$ is the CMB power spectrum, $w_{\ell}$ contains the
instrumental and pixelization beams, $C^{noise}_{\ell}$ is the power
spectrum of the noise, $\Delta_{\ell_1 \ell_2 \ell_3} = 1 +
2\delta_{\ell_1 \ell_2}\delta_{\ell_2 \ell_3} + \delta_{\ell_1 \ell_2}
+ \delta_{\ell_2 \ell_3} + \delta_{\ell_1 \ell_3}$. The error-bar of
the $f_{nl}$ parameter is given by \citep{komatsu2001}:
\begin{equation}
\sigma^{-2}\big(f_{nl}\big) = \sum_{\ell_1 \le \ell_2 \le \ell_3}
\frac{\big(B^{prim}_{\ell_1 \ell_2 \ell_3}\big)^2}
     {\big(\Delta_{\ell_1 \ell_2 \ell_3} C^{tot}_{\ell_1}
       C^{tot}_{\ell_2} C^{tot}_{\ell_3}\big)}.
\label{eq:fnl_error_point_sources}
\end{equation}
This estimator can be generalised to include the polarization:
\begin{align}
\nonumber
\Delta f_{nl}  &=   \sigma^2\big(f_{nl}\big) \times   \\
\sum_{ijkrst}\sum_{\ell_1 \le \ell_2 \le \ell_3 \le \ell_{max}}&
\big(B_{\ell_1 \ell_2 \ell_3}^{ps}\big)^{ijk} C_{ijk,rst}^{-1}\big(B_{\ell_1 \ell_2 \ell_3}^{prim}\big)^{rst},
\label{eq:fnl_bias_point_sources_pol}
\end{align}
where
\begin{align}
\nonumber
\sigma^{-2}\big(f_{nl}\big) & =  \\
\sum_{ijkrst}\sum_{\ell_1 \le \ell_2 \le \ell_3 \le \ell_{max}} &
\big(B_{\ell_1 \ell_2 \ell_3}^{prim}\big)^{ijk} C_{ijk,rst}^{-1}\big(B_{\ell_1 \ell_2 \ell_3}^{prim}\big)^{rst}.
\label{eq:fnl_error_point_sources_pol}
\end{align}
The covariance matrix $C_{ijk,rst}$ is given by \citep{yadav2007}:
\begin{align}
\nonumber
C_{ijk,prq} & = C^{ip}_{\ell_1}C^{jq}_{\ell_2} C^{kr}_{\ell_3}+C^{ip}_{\ell_1}C^{jr}_{\ell_2} C^{kq}_{\ell_3}\delta_{\ell_2 \ell_3}+C^{iq}_{\ell_1}C^{jp}_{\ell_2} C^{kr}_{\ell_3}\delta_{\ell_1 \ell_2}\\
\nonumber
& +C^{ir}_{\ell_1}C^{jq}_{\ell_2} C^{kp}_{\ell_3}\delta_{\ell_1 \ell_3}+C^{iq}_{\ell_1}C^{jr}_{\ell_2} C^{kp}_{\ell_3}\delta_{\ell_1 \ell_2}\delta_{\ell_2 \ell_3}\delta_{\ell_3 \ell_1}\\
& +C^{ir}_{\ell_1}C^{jp}_{\ell_2} C^{kq}_{\ell_3}\delta_{\ell_1 \ell_3}\delta_{\ell_2 \ell_1}\delta_{\ell_2 \ell_3}.
\label{eq:covariance_bispectrum_pol}
\end{align}
The indices $\{i,j,k,r,s,t \}$ cover the intensity $T$ and the
polarization $E$ and $B$ modes. The cosmological signal for the $B$
mode is negligible. Including the point source $B$ mode signal in the
estimator does not improve the estimates of $\sigma(f_{nl})$ and
$\Delta(f_{nl})$. Therefore we just include in the estimator the T and
E modes for the case with polarization. As in the case with intensity
only, each of the power or cross-power spectrum in
Eq. (\ref{eq:covariance_bispectrum_pol}) contain the CMB signal,
instrumental beam, instrumental noise spectrum and the point source
spectrum contributions.
\section{Results}
\label{sect:results}
\begin{table}
  \center
  \caption{\label{table_instr} Instrumental properties for a possible
    future COrE-like mission, based on values given by
    \citet{core2011}.}
\begin{tabular}{c|cccccc}
\hline
$\nu$ (GHz) & 45 & 75 & 105 & 165 & 225 & 375 \\
FWHM (arcmin) & 23.3 & 14.0 & 10.0 & 6.4 & 4.7 & 2.8\\
T RMS noise$^*$  & 5.25 & 2.73 & 2.68 & 2.67 & 2.64 & 68.6\\
Pol. RMS noise$^*$  & 9.07 & 4.72 & 4.63 & 4.61 & 4.57 & 119\\
\hline
\end{tabular}
\begin{tablenotes}
\small
\item $^*$ $\mu\,K$\,arcmin
\end{tablenotes}
\end{table}
\begin{table*}
\begin{center}
\caption{The expected uncertainty $\sigma(f_{nl})$, bias $\Delta f_{nl}$
  and relative bias $\Delta f_{nl}/\sigma(f_{nl})$ at
  $\ell_{max}=2048$ from temperature only and temperature plus polarization
  for an ideal mission. $S_c=$ 0.3 Jy is adopted.
  \label{table_ideal_bias_0d3}}
\begin{tabular}{c|c||cccccc|c|cccccc|}
\hline
& case & T & T &  T & T & T & T & & T+E  & T+E  & T+E &  T+E & T+E & T+E \\
& Freq. & 45 & 75 & 105 & 165 & 225 & 375 & & 45 & 75 & 105 & 165 & 225 & 375\\
\hline
& $\sigma(f_{nl})$ &7.4 &        4.9 &        4.4 &        4.2 &        5.2  &       29.7   &&    2.8 &        2.3 &        2.1 &        2.1 &        2.3  &        3.3 \\
local & $\Delta f_{nl}$ &82.1 &       10.3 &        2.1 &        0.4 &        0.7  &       25.6   &&   -9.7 &       -2.6 &       -0.6 &       -0.1 &       -0.1  &       -0.1 \\
& $\Delta f_{nl}/\sigma(f_{nl})$ &  11.1 &        2.1 &        0.5 &        0.1 &        0.1  &        0.9    &&  -3.5 &       -1.1 &       -0.3 &       -0.1 &       -0.1  &       -0.0 \\
\hline
& $\sigma(f_{nl})$ &75.4 &       60.4 &       56.8 &       55.9 &       65.1  &      232.4   &&   24.0 &       20.3 &       19.5 &       19.2 &       21.1  &       35.3 \\
equilateral &$\Delta f_{nl}$ & 4629.3 &      599.9 &      107.2 &       17.2 &       49.1  &      950.7   &&  480.2 &       54.5 &        7.3 &        0.8 &        4.0  &       17.9 \\
& $\Delta f_{nl}/\sigma(f_{nl})$ & 61.4 &        9.9 &        1.9 &        0.3 &        0.8  &        4.1  &&    20.0 &        2.7 &        0.4 &        0.0 &        0.2  &        0.5 \\
\hline
& $\sigma(f_{nl})$ &40.1 &       29.7 &       27.4 &       26.8 &       32.0  &      139.0   &&   11.6 &        9.8 &        9.4 &        9.2 &       10.1  &       16.5 \\
orthogonal & $\Delta f_{nl}$ & 955.5 &       51.7 &        2.5 &       -0.2 &        9.7  &      371.4   &&  308.7 &       48.3 &        9.4 &        1.6 &        3.4  &        7.2 \\
& $\Delta f_{nl}/\sigma(f_{nl})$ &23.8 &        1.7 &        0.1 &       -0.0 &        0.3  &        2.7    &&  26.7 &        4.9 &        1.0 &        0.2 &        0.3  &        0.4 \\
\hline
& $\sigma(f_{nl})$ &73.7 &       53.3 &       49.0 &       47.9 &       57.6  &      284.7  &&    20.6 &       17.4 &       16.6 &       16.4 &       17.9  &       29.1 \\
flat & $\Delta f_{nl}$ & 598.6 &      150.5 &       36.0 &        6.7 &        3.4  &      -66.0  &&  -311.6 &      -56.3 &      -12.2 &       -2.2 &       -3.9  &       -5.2 \\
& $\Delta f_{nl}/\sigma(f_{nl})$ &8.1 &        2.8 &        0.7 &        0.1 &        0.1  &       -0.2  &&   -15.1 &       -3.2 &       -0.7 &       -0.1 &       -0.2  &       -0.2 \\
\hline
\end{tabular}
\caption{The expected uncertainty $\sigma(f_{nl})$, bias $\Delta f_{nl}$
  and relative bias $\Delta f_{nl}/\sigma(f_{nl})$ at
  $\ell_{max}=2048$ from temperature only and temperature plus polarization
  for an ideal mission. $S_c=$ 1 Jy is adopted.
  \label{table_ideal_bias_1d0}}
\begin{tabular}{c|c||cccccc|c|cccccc|}
\hline
& case & T & T &  T & T & T & T & & T+E  & T+E  & T+E &  T+E & T+E & T+E \\
& Freq. & 45 & 75 & 105 & 165 & 225 & 375 & & 45 & 75 & 105 & 165 & 225 & 375\\
\hline
& $\sigma(f_{nl})$ &9.9 &        5.9 &        4.9 &        4.4 &        5.3  &       29.8   &&    3.2 &        2.5 &        2.3 &        2.1 &        2.3  &        3.3 \\
local & $\Delta f_{nl}$ &351.1 &       68.5 &       17.6 &        3.4 &        2.3  &       25.6   &&   -5.9 &      -12.6 &       -4.4 &       -1.0 &       -0.5  &       -0.1 \\
& $\Delta f_{nl}/\sigma(f_{nl})$ &35.6 &       11.6 &        3.6 &        0.8 &        0.4  &        0.9   &&   -1.9 &       -5.1 &       -2.0 &       -0.5 &       -0.2  &       -0.0 \\
\hline
& $\sigma(f_{nl})$ &88.7 &       66.6 &       59.9 &       57.0 &       65.6  &      229.8  &&    27.7 &       21.8 &       20.2 &       19.5 &       21.2  &       35.3 \\
equilateral & $\Delta f_{nl}$ &17689.6 &     4075.1 &     1017.5 &      174.2 &      135.5  &      948.9  &&  2226.3 &      390.7 &       90.7 &       11.5 &       10.8  &       20.5 \\
& $\Delta f_{nl}/\sigma(f_{nl})$ &199.4 &       61.2 &       17.0 &        3.1 &        2.1  &        4.1   &&   80.4 &       17.9 &        4.5 &        0.6 &        0.5  &        0.6 \\
\hline
& $\sigma(f_{nl})$ &50.2 &       33.8 &       29.3 &       27.5 &       32.3  &      138.6   &&   13.4 &       10.5 &        9.7 &        9.4 &       10.1  &       16.5 \\
orthogonal & $\Delta f_{nl}$ & 5411.3 &      573.7 &       80.9 &        3.6 &       16.4  &      377.8 &&   1007.2 &      297.6 &       82.1 &       15.1 &        9.3  &        7.1 \\
& $\Delta f_{nl}/\sigma(f_{nl})$ &107.8 &       17.0 &        2.8 &        0.1 &        0.5  &        2.7   &&   75.3 &       28.4 &        8.5 &        1.6 &        0.9  &        0.4 \\
\hline
& $\sigma(f_{nl})$ &94.6 &       61.3 &       52.7 &       49.2 &       58.2  &      284.5   &&   23.8 &       18.6 &       17.3 &       16.6 &       18.0  &       29.2 \\
flat & $\Delta f_{nl}$ & 455.7 &      784.2 &      263.4 &       59.1 &       26.6  &      -69.0  &&  -770.4 &     -326.8 &      -96.4 &      -19.6 &      -10.8  &       -4.1 \\
& $\Delta f_{nl}/\sigma(f_{nl})$ &4.8 &       12.8 &        5.0 &        1.2 &        0.5  &       -0.2  &&   -32.4 &      -17.5 &       -5.6 &       -1.2 &       -0.6  &       -0.1 \\
\hline
\end{tabular}
\end{center}
\end{table*}
We have studied the impact of unresolved point sources on the possible
measurement of the $f_{nl}$ parameter for two cases: $a$) an ideal
experiment without instrumental beam nor instrumental noise; $b$) a
possible future COrE-like mission, sensitive to temperature and
polarization, with instrumental parameters adopted from
\citet{core2011} and summarised in Table\,\ref{table_instr}. The CMB
power spectrum and transfer function for temperature and polarization
needed to estimate the $f_{nl}$ bias and its uncertainty
(Eqs. \ref{eq:fnl_bias_point_sources},
\ref{eq:fnl_error_point_sources}, \ref{eq:fnl_bias_point_sources_pol}
and \ref{eq:fnl_error_point_sources_pol}) are numerically computed
with CAMB \citep{lewis2000} using the cosmological parameters that
best-fits WMAP 7-yr data. In this section we present results for the
uncertainty on the $f_{nl}$ parameter, $\sigma(f_{nl})$, for the
expected bias, $\Delta(f_{nl})$, and for the relative bias,
$\Delta(f_{nl})/\sigma(f_{nl})$ (see
Fig.\,\ref{fig:error_fnl_ideal_t_0d3},
Tables\,\ref{table_ideal_bias_0d3} and \ref{table_ideal_bias_1d0} for
the ideal case, and Fig.\,\ref{fig:error_fnl_t_0d3},
Tables\,\ref{table_core_bias_0d3} and \ref{table_core_bias_1d0} for a
COrE--like mission). We compare results obtained with temperature only
and with temperature plus polarization. Four different $f_{nl}$ shapes
are taken into account, i.e. local, equilateral, orthogonal and
flat. Fig.\,\ref{fig:error_fnl_ideal_t_0d3} and
\ref{fig:error_fnl_t_0d3} are for $S_c=0.3$\,Jy and for frequencies
75, 105, 165 and 225\,GHz. In the Tables we also provide the results,
evaluated at $\ell_{max}=2048$, for $S_c=1.0$\,Jy and for 45 and
375\,GHz.

The following results assume full sky. This is an unrealistic hypothesis 
as Galactic Plane and strong point sources have to be masked. How
$\sigma(f_{nl})$ depends on the mask and on the fraction of the sky
available, $f_{sky}$, is something complex and outside the purpose of
this work. However, for large areas of the sky ($f_{sky}\ga0.7$) the
following approximation can be used \citep{komatsu2002}:
\begin{equation}
  \sigma(f_{nl})_{\rm cutsky} = \sigma(f_{nl})_{\rm fullsky}/\sqrt{f_{sky}}\,.
\end{equation}
The bias on $f_{nl}$ does not depend on $f_{sky}$, as it can be seen
from Eq.\,\ref{eq:fnl_bias_point_sources}.
\subsection{Uncertainty}
In Eqs.\,\ref{eq:fnl_error_point_sources} and
\ref{eq:fnl_error_point_sources_pol} of the previous section we have
provided the analytical formulas to estimate the uncertainty on the
$f_{nl}$ parameter. As we can see, the point sources contribution
enters only in the total angular power spectrum
(Eq.\,\ref{total_cls}), in the sum $C_{\ell}^{CMB}+C_{\ell}^{ps}$. In
the frequency range 75--225\,GHz, $C_{\ell}^{ps}$ is always
subdominant with respect to T and E--mode CMB spectra up to
$\ell\sim2000$ (see Fig.\,\ref{f2}). Therefore, as illustrated by
Fig.\,\ref{fig:error_fnl_ideal_t_0d3} and \ref{fig:error_fnl_t_0d3},
the impact of point sources on the uncertainty in the determination of
the $f_{nl}$ parameter is small or negligible, especially when
polarization is considered. In this case, in fact, $\sigma(f_{nl})$ is
almost indistinguishable with the intrinsic $f_{nl}$ uncertainty due
to the cosmic variance (black curves in the plots, indicated as ``no
sources''). This is true also for the ``temperature only'' case at 105
and 165\,GHz, i.e. the frequencies where the signal from point sources
is minimum. At 75\,GHz the uncertainty becomes slightly higher only at
$\ell\ga1500$ because of the increasing contribution of ERS; while, a
larger $\sigma(f_{nl})$ is obtained at 225\,GHz in the all range of
$\ell$s, clearly due to the clustered signal of IR sources. At
353\,GHz the point sources contamination grows dramatically, and the
$\sigma(f_{nl})$ is around 4--6 times higher than at 225\,GHz. On the
other hand, the loss of sensitivity between 75 and 45\,GHz is not so
strong, due to the flatter spectral behaviour of ERS. These results
seem to be quite independent of the two flux density limits considered, even
at 45 and 75\,GHz, where $\sigma(f_{nl})$ is about 10--20\% larger
(see Table\,\ref{table_ideal_bias_0d3}--\ref{table_core_bias_1d0}).

When polarization is taken into account, $f_{nl}$ error--bars are
reduced approximately by a factor of 2--3 accordingly to the shape, in
the frequency range 75--225\,GHz. At 45 and 375\,GHz, polarization
permits to achieve a very large improvement in the $f_{nl}$
error--bars, but only for the ideal case. In a COrE--like mission the
instrumental noise in polarization saturates the $f_{nl}$ error--bars
at these frequencies and prevents to reduce them as in the ideal case.

The values of $\sigma(f_{nl})$ obtained from a COrE--like mission are
very close to the results of the ideal case. This means that
observational performances, in terms of instrumental noise and beam,
given in Table\,\ref{table_instr}, are almost optimal for $f_{nl}$
measurements, at least up to $\ell\sim2000$. Only in the 75--GHz
channel (and partially in the 105--GHz channel) the effect of the beam
is observed in Fig.\,\ref{fig:error_fnl_t_0d3}, where $\sigma(f_{nl})$
starts to be constant at $\ell_{max}\ga1500$. In the 165-- and
225--GHz channels the $f_{nl}$ uncertainty is always decreasing up to
$\ell_{max}=2048$. Note however that this is not the case in a
Planck-like mission, where the noise contamination in the E mode is
more important and the improvement from T only to T and polarization
is 20-30\% \citep{martinez2012}.
\subsection{Bias}
\begin{figure*}
\includegraphics[width=8.0cm, height=5.5cm, angle= 0]{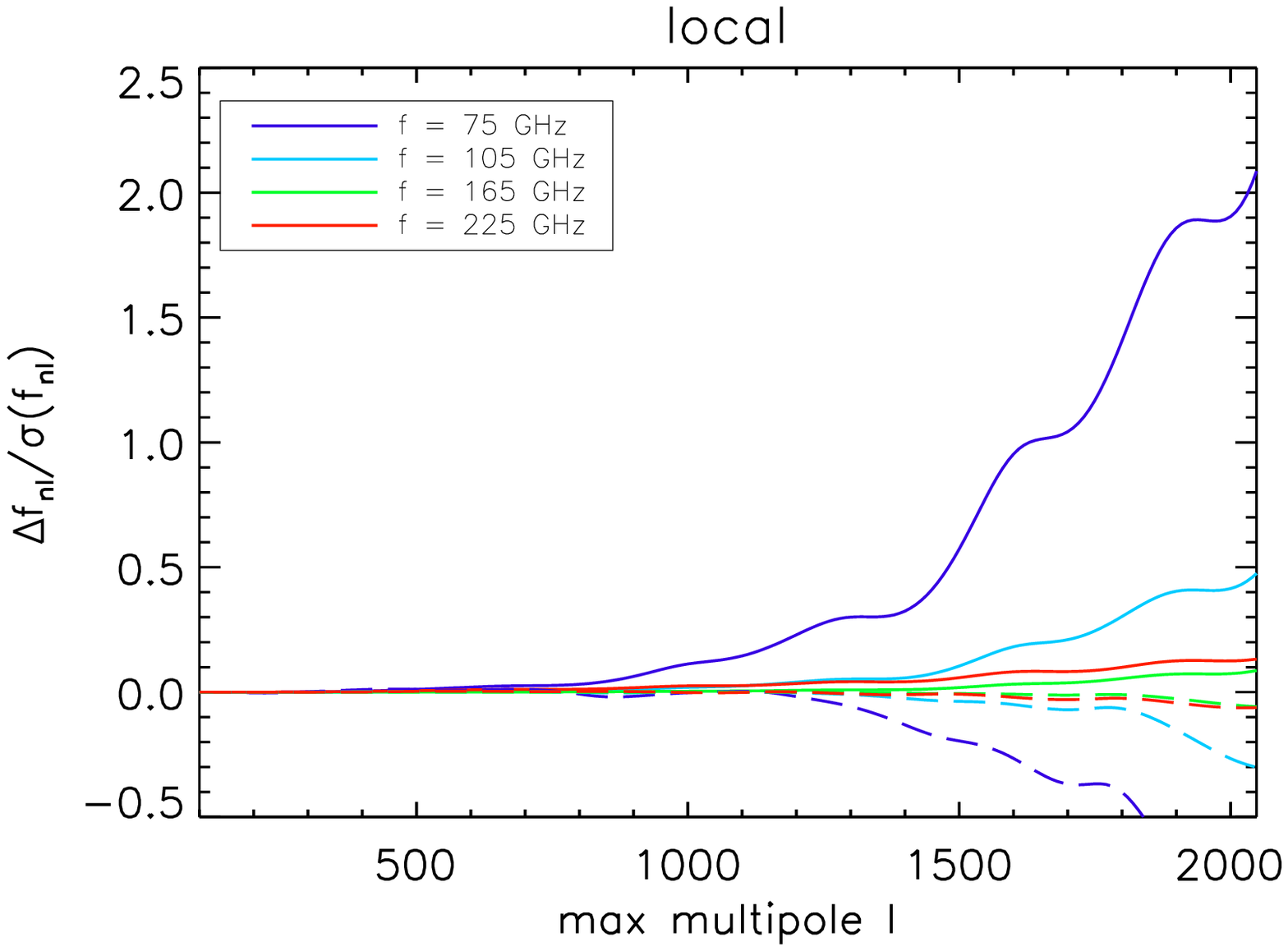}
\includegraphics[width=8.0cm, height=5.5cm, angle= 0]{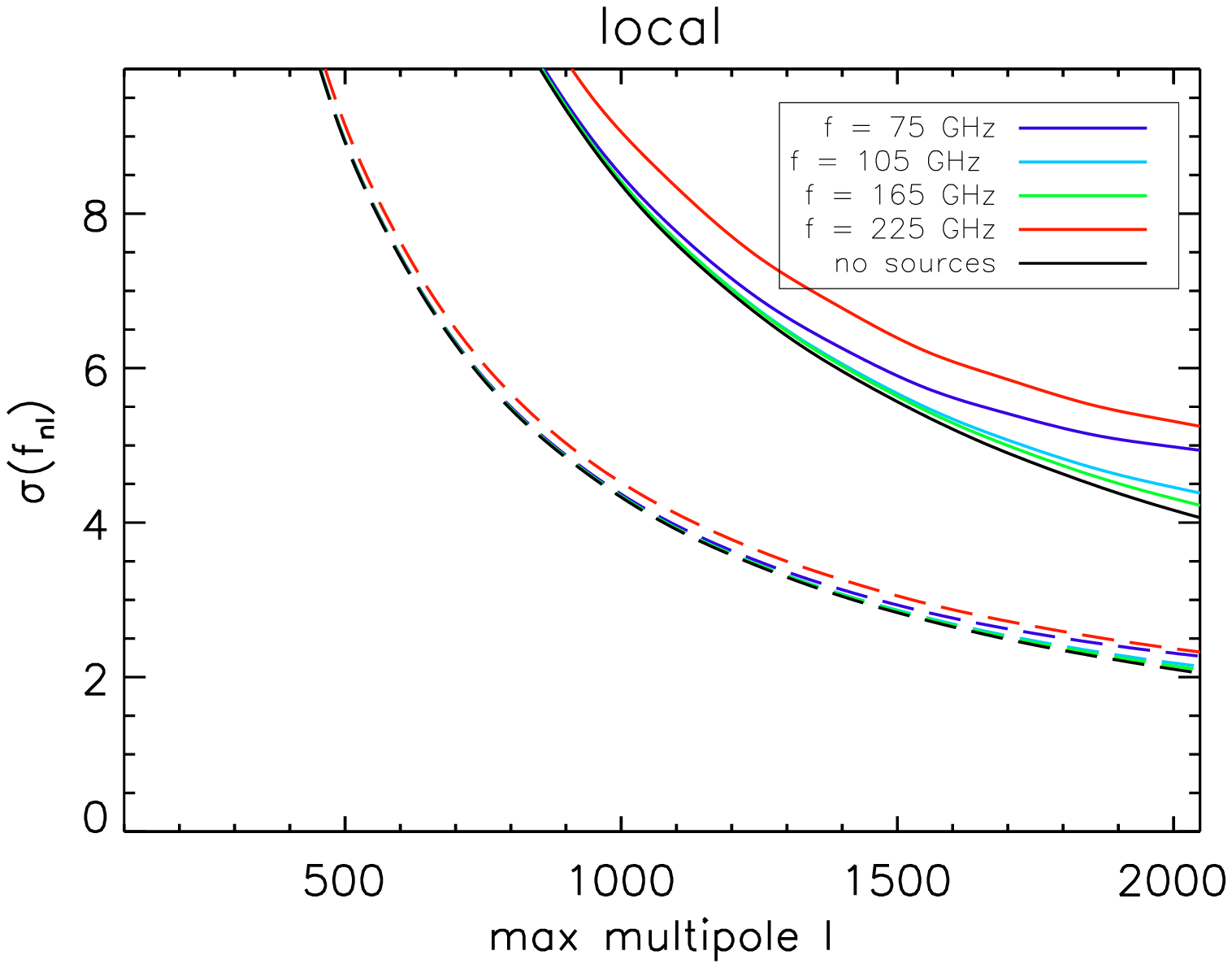}
\includegraphics[width=8.0cm, height=5.5cm, angle= 0]{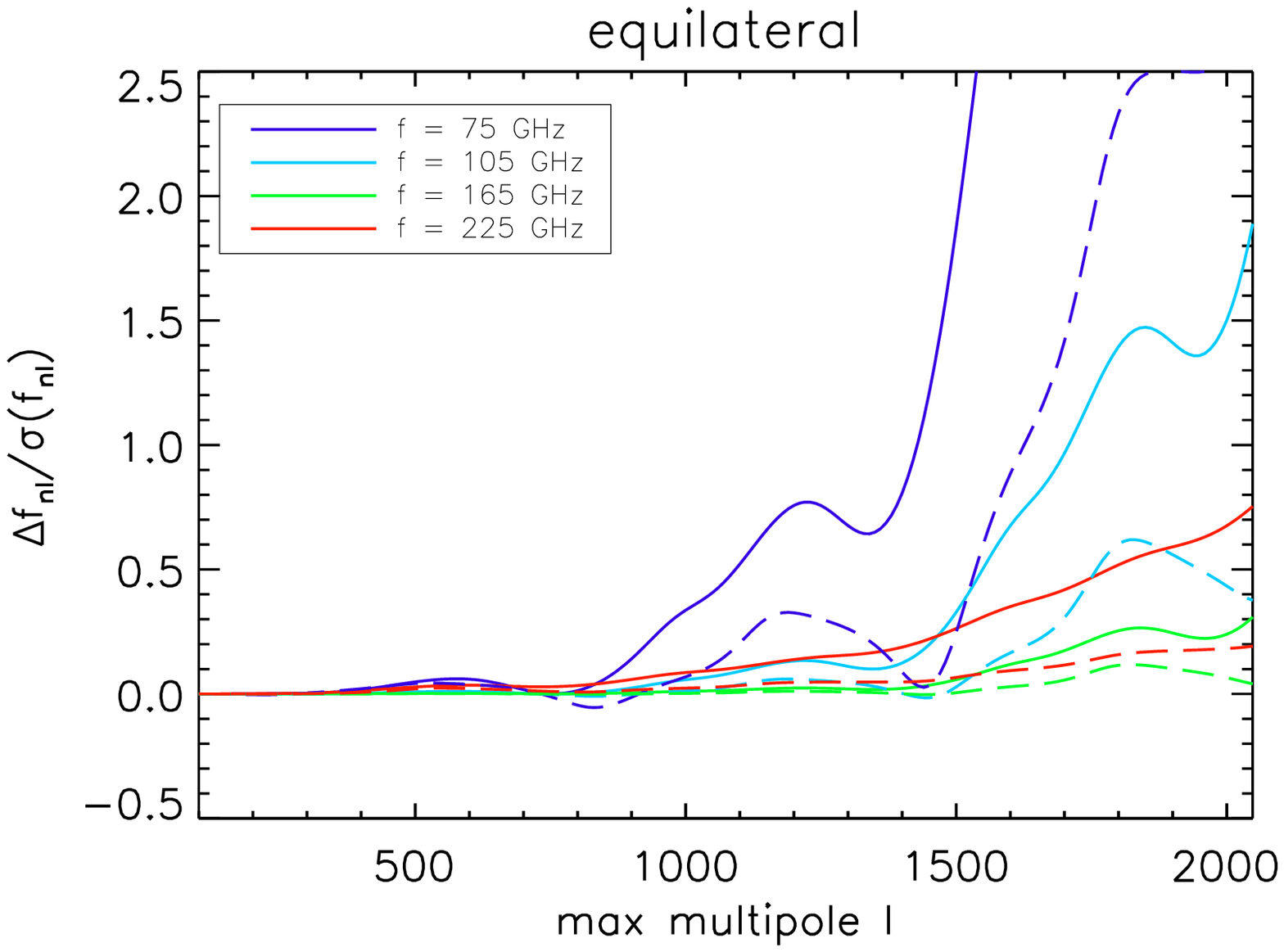}
\includegraphics[width=8.0cm, height=5.5cm, angle= 0]{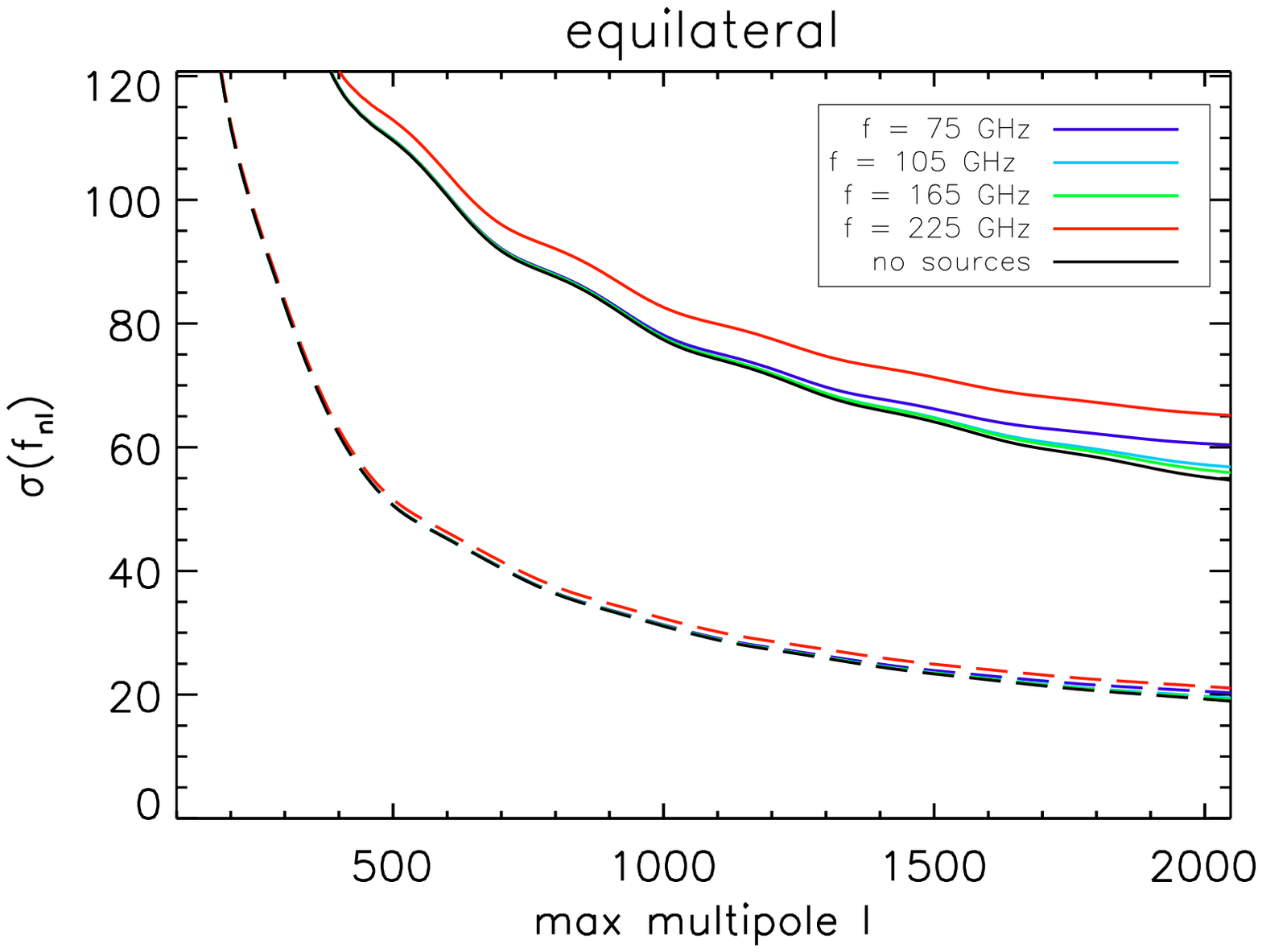}
\includegraphics[width=8.0cm, height=5.5cm, angle= 0]{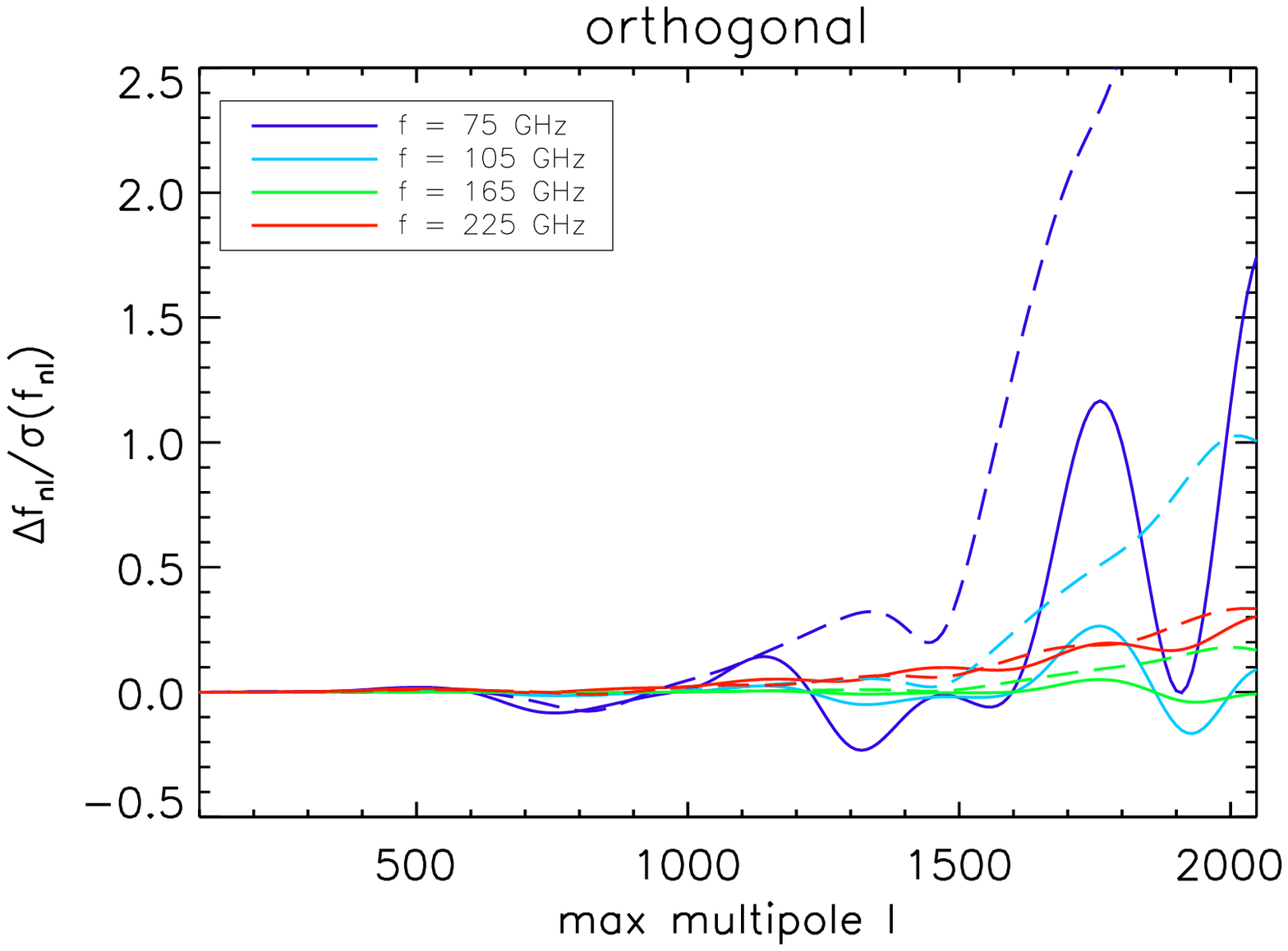}
\includegraphics[width=8.0cm, height=5.5cm, angle= 0]{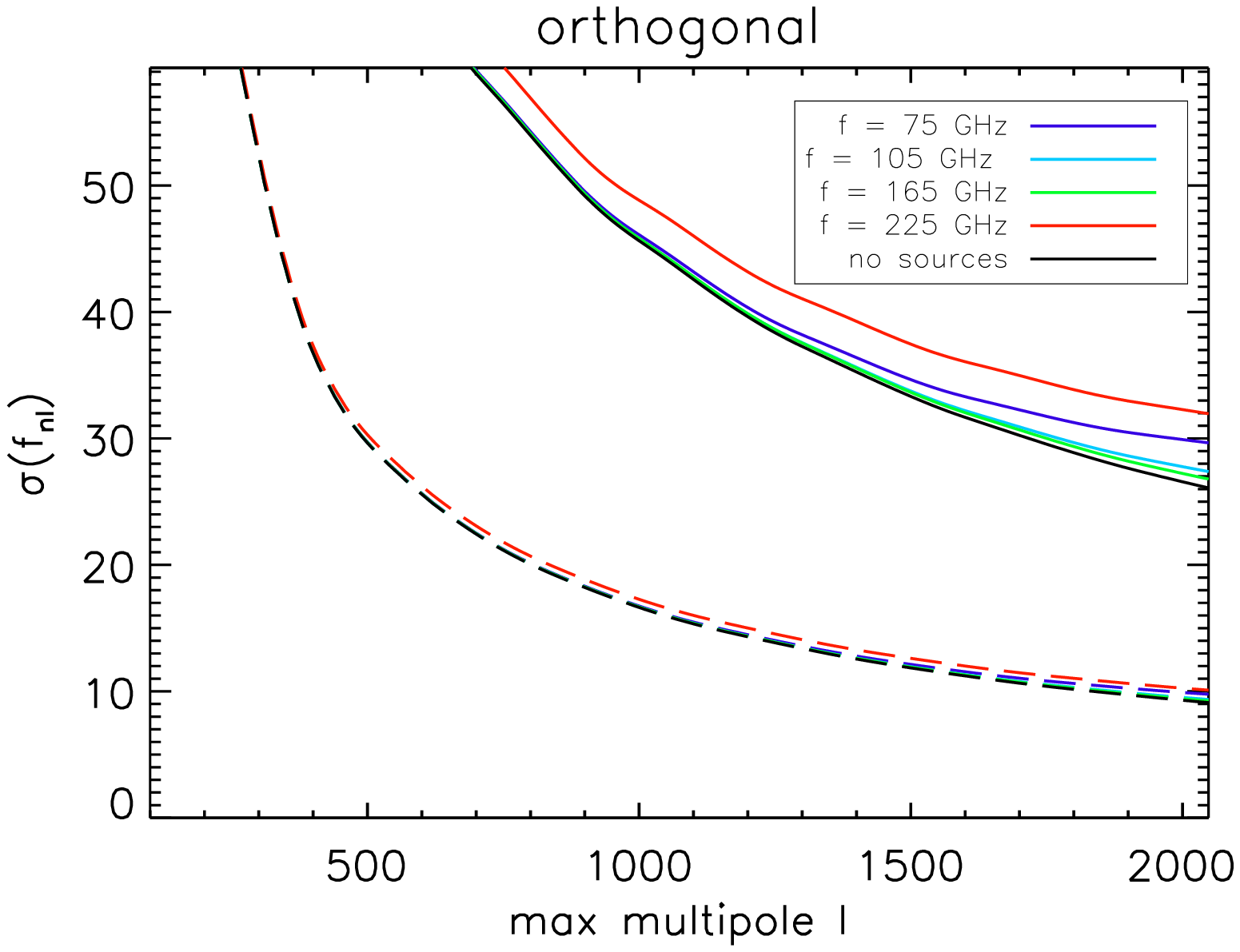}
\includegraphics[width=8.0cm, height=5.5cm, angle= 0]{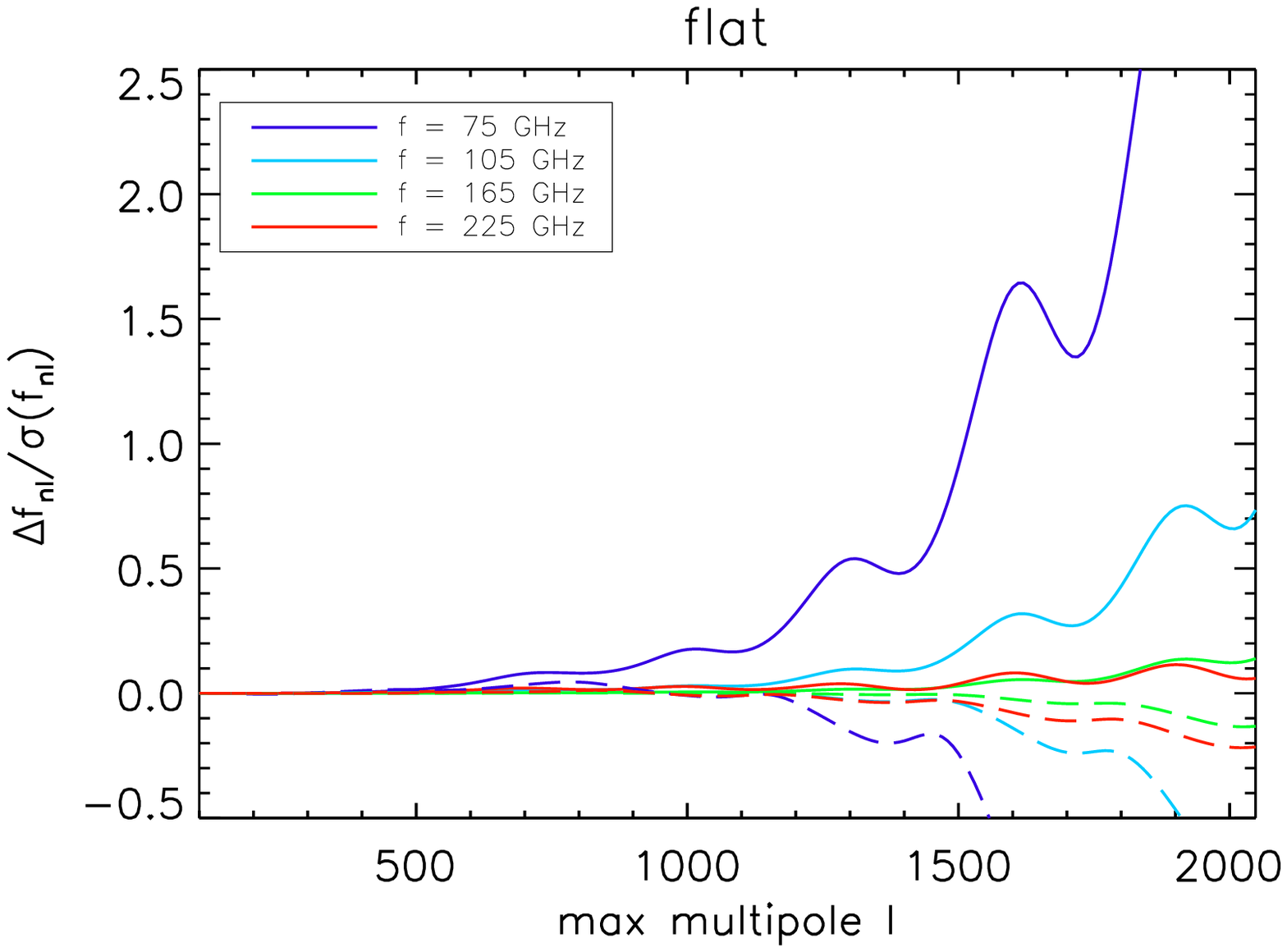}
\includegraphics[width=8.0cm, height=5.5cm, angle= 0]{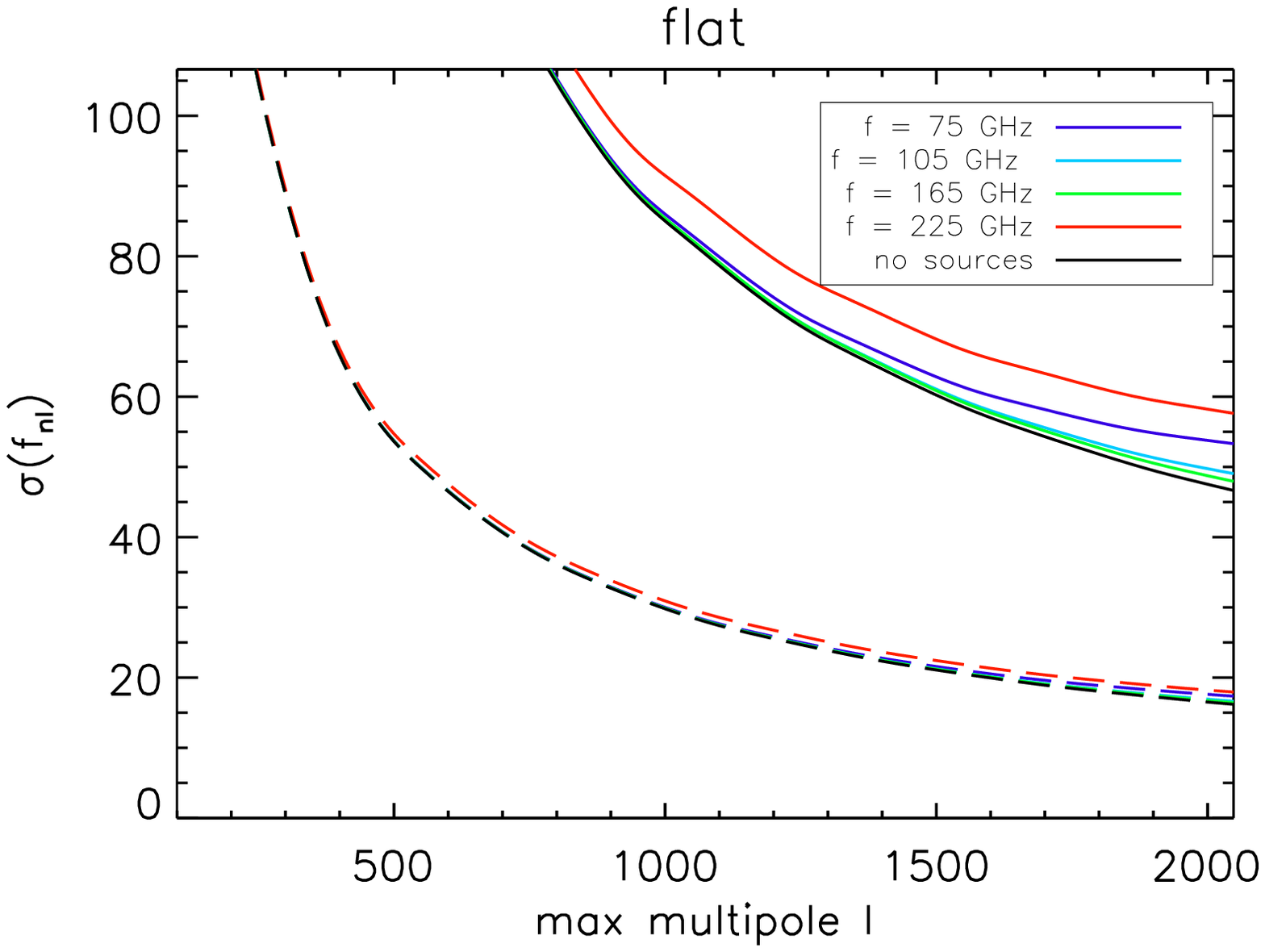}
\caption{\label{fig:error_fnl_ideal_t_0d3} {\it From left to right}:
  the relative bias $\Delta(f_{nl})/\sigma(f_{nl})$ and the $f_{nl}$
  uncertainty $\sigma(f_{nl})$ produced by unresolved point sources as
  a function of $\ell_{max}$ for the ideal case. We consider the flux density
  limit $S_c=$0.3\,Jy and the following frequencies: 75, 105, 165, 225
  GHz and the configuration without point sources (black line in right
  panels). From top to bottom, we plot results for the local,
  equilateral, orthogonal and flat $f_{nl}$ shapes. The solid lines
  take into account only temperature, while the long dashed lines
  include temperature and polarization.}
\end{figure*}

\begin{figure*}
\includegraphics[width=8.0cm, height=5.5cm, angle= 0]{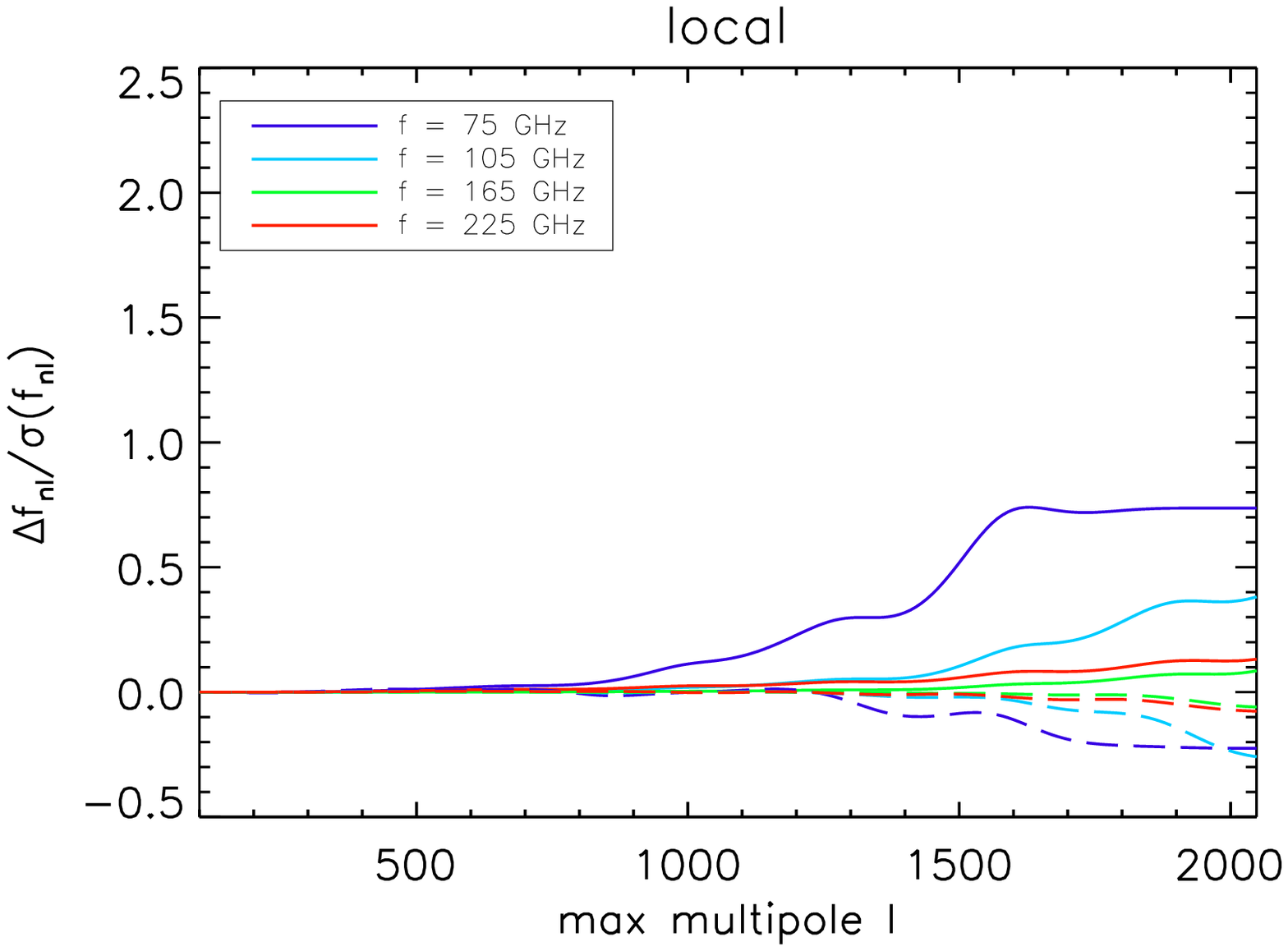}
\includegraphics[width=8.0cm, height=5.5cm, angle= 0]{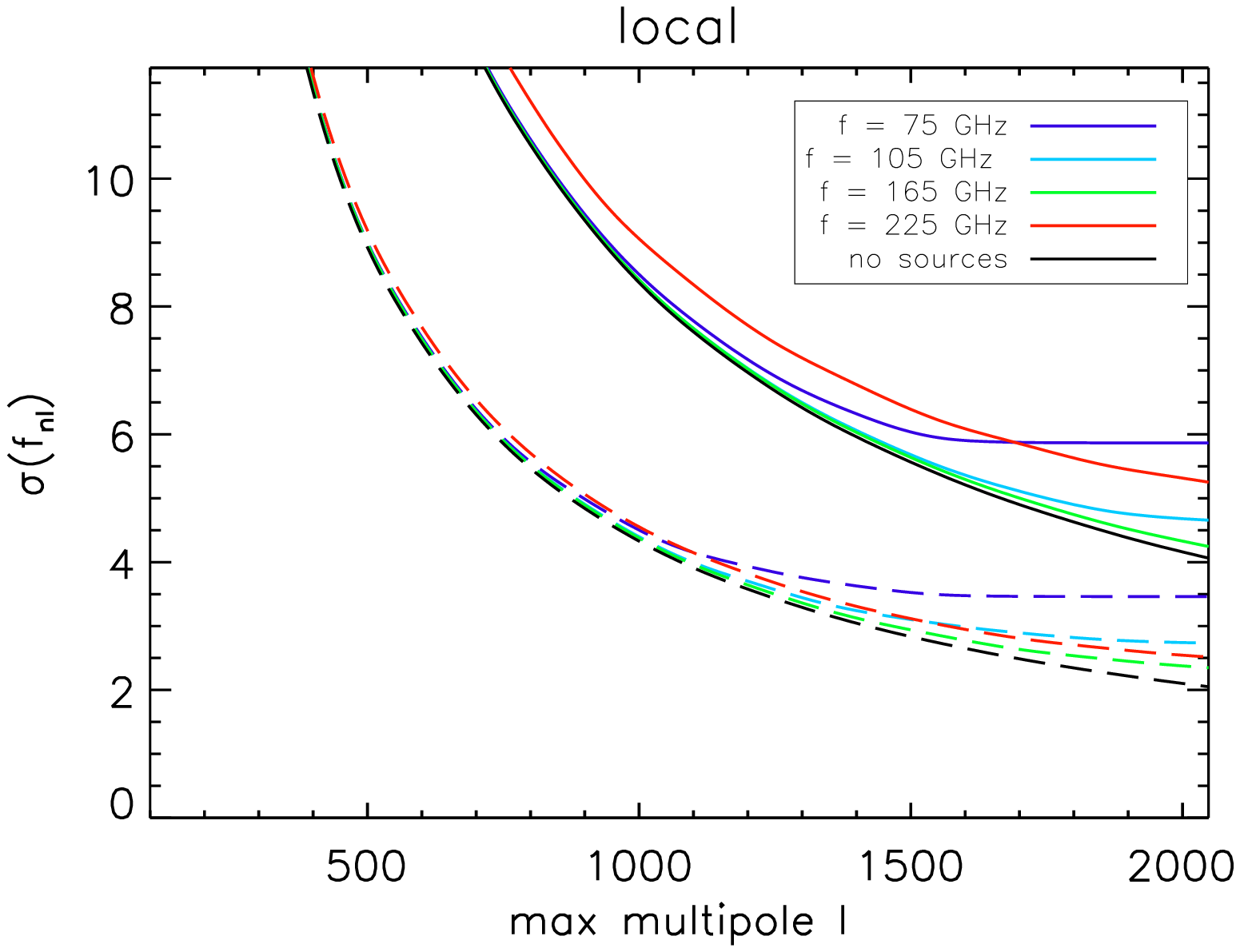}
\includegraphics[width=8.0cm, height=5.5cm, angle= 0]{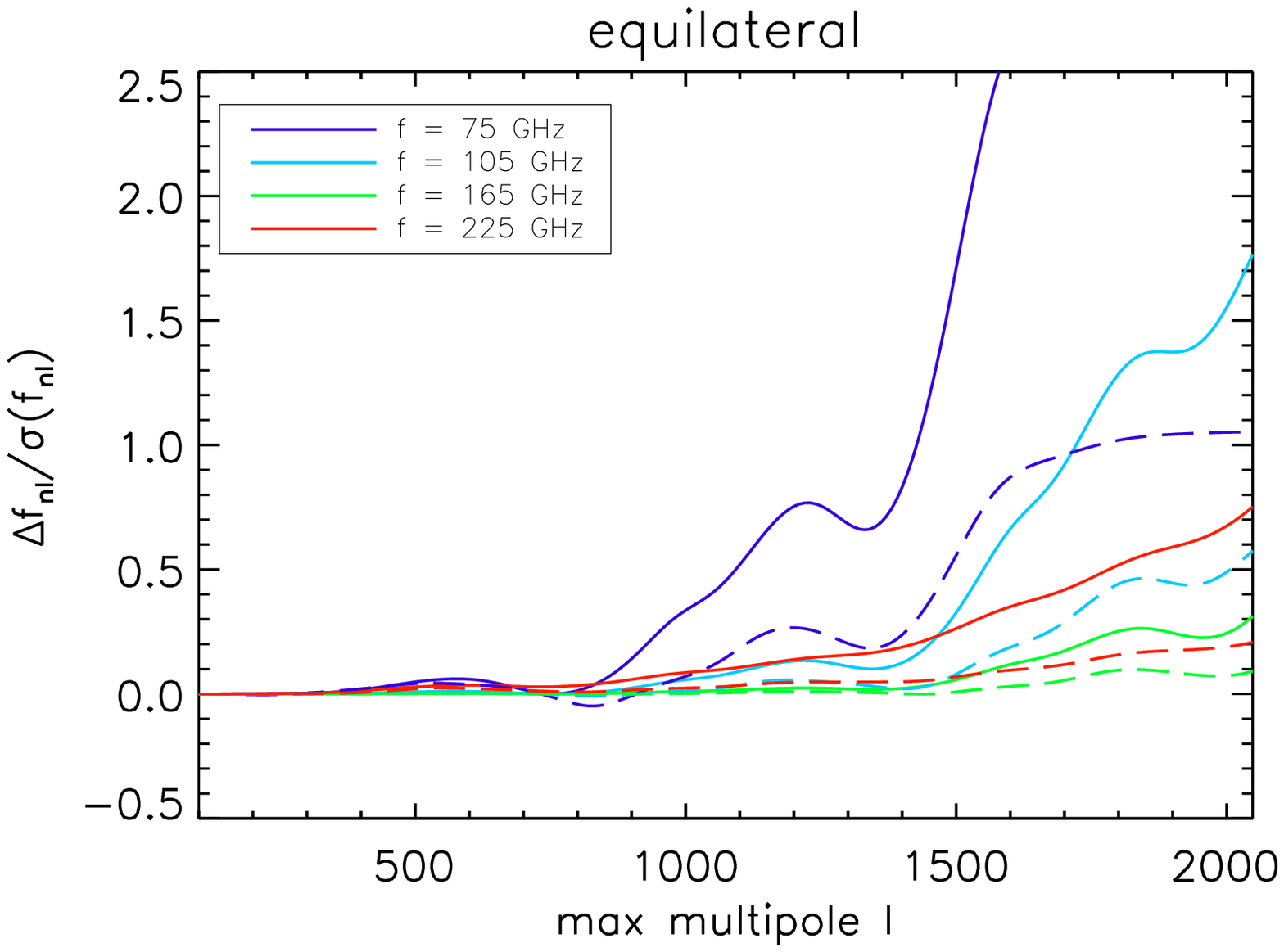}
\includegraphics[width=8.0cm, height=5.5cm, angle= 0]{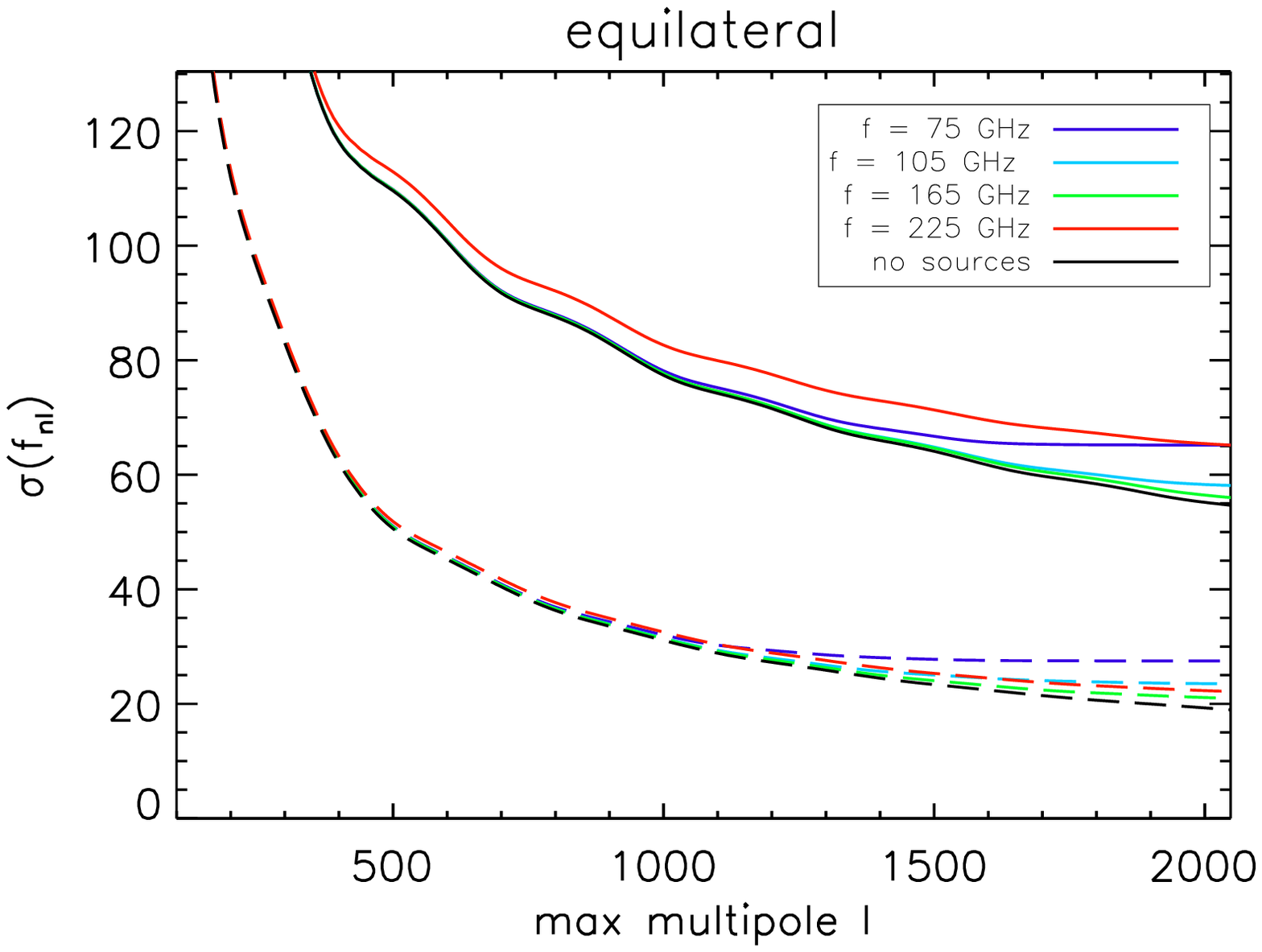}
\includegraphics[width=8.0cm, height=5.5cm, angle= 0]{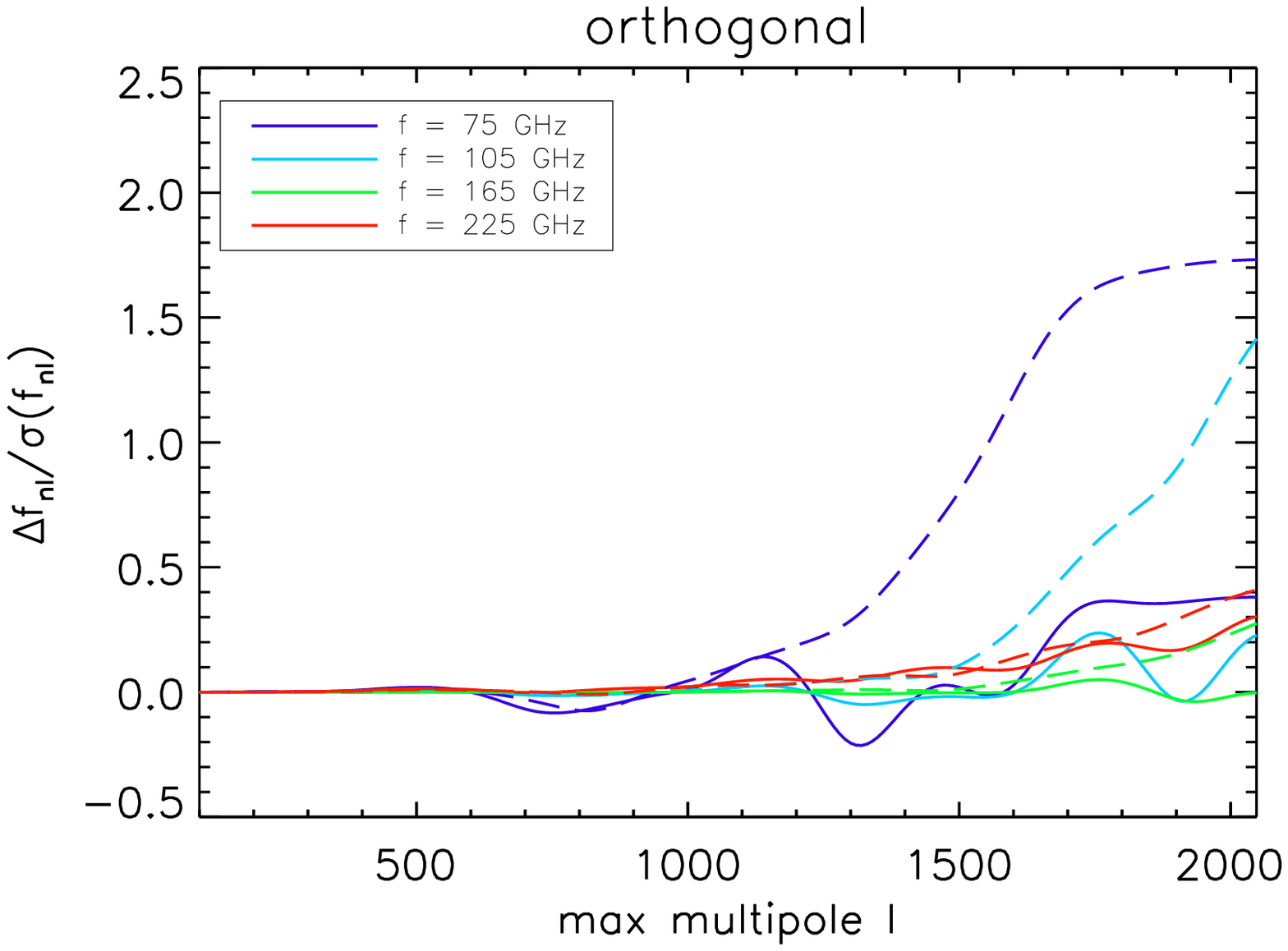}
\includegraphics[width=8.0cm, height=5.5cm, angle= 0]{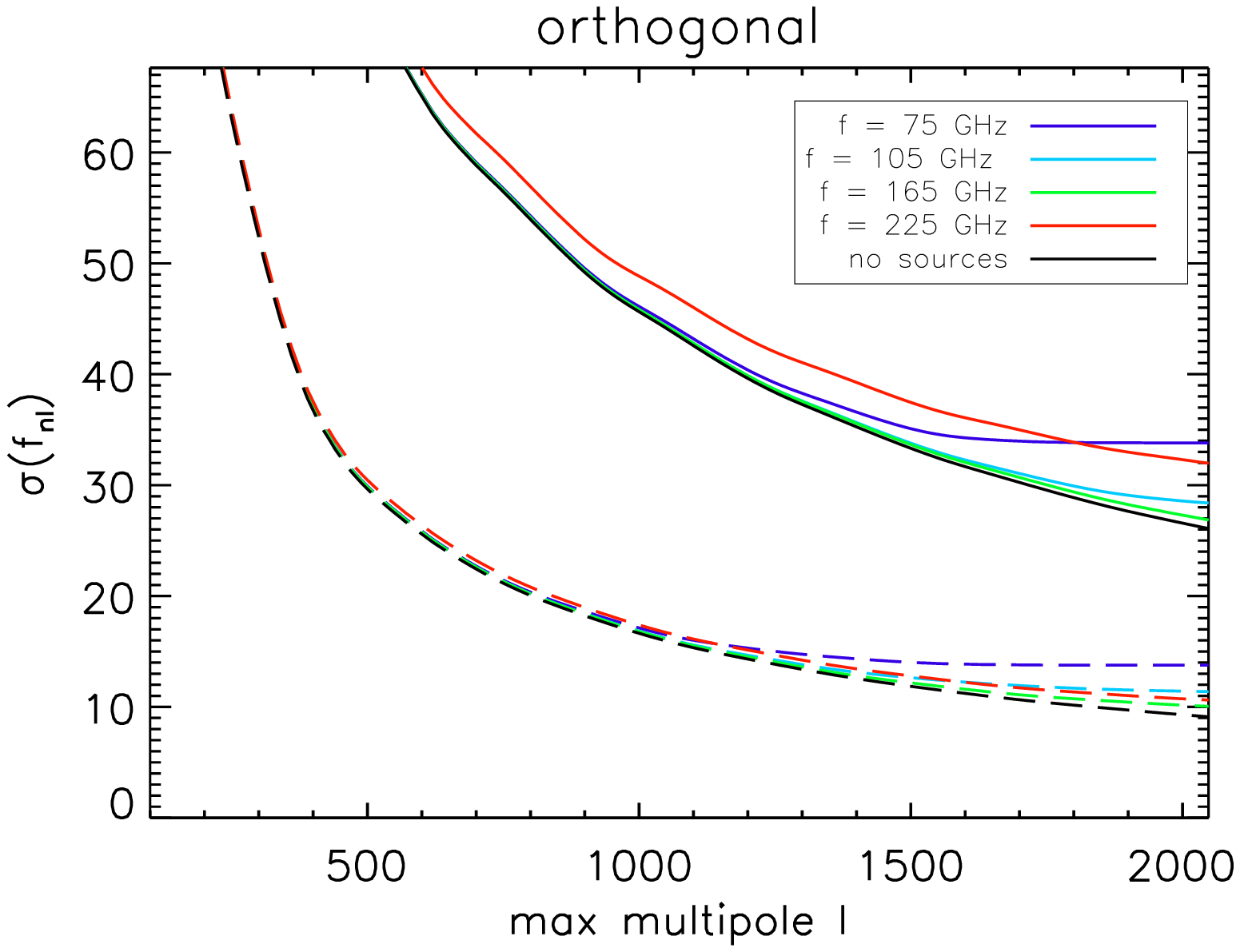}
\includegraphics[width=8.0cm, height=5.5cm, angle= 0]{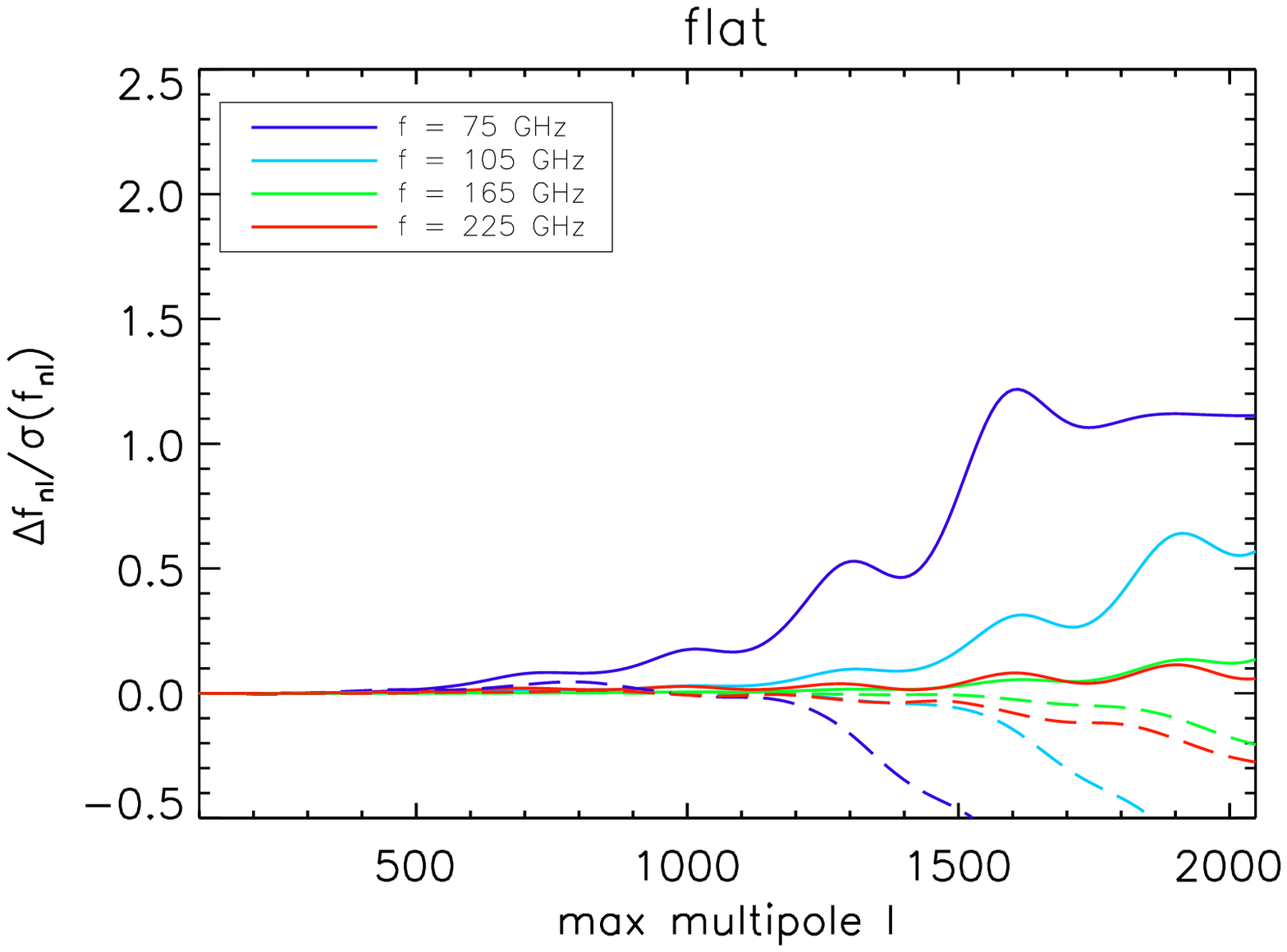}
\includegraphics[width=8.0cm, height=5.5cm, angle= 0]{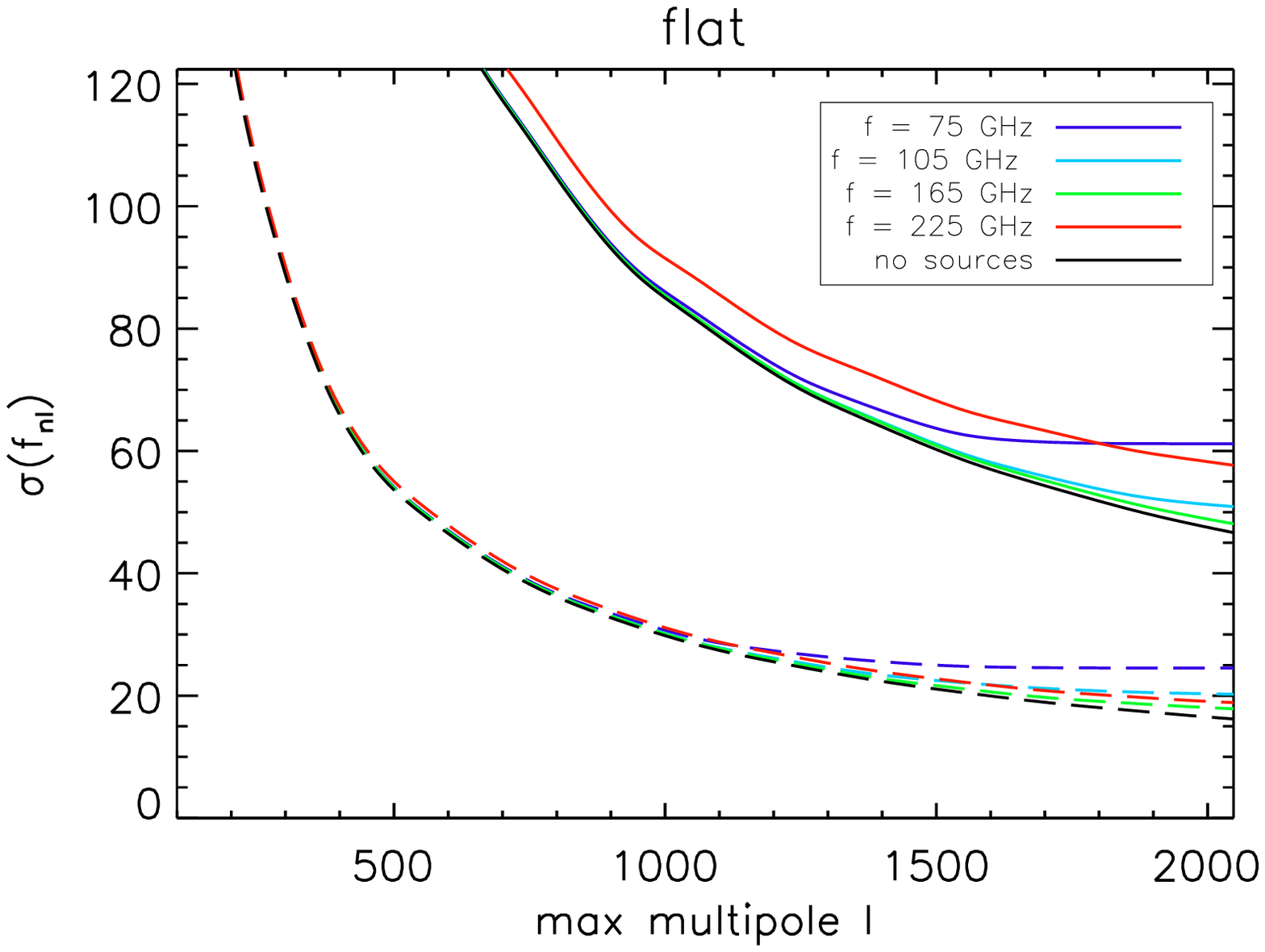}
\caption{\label{fig:error_fnl_t_0d3} As in
  Fig.\,\ref{fig:error_fnl_ideal_t_0d3} but for a COrE-like mission.}
\end{figure*}

As expected, the bias (or the relative bias, shown in
Fig. \,\ref{fig:error_fnl_ideal_t_0d3} and \ref{fig:error_fnl_t_0d3})
rapidly increases with $\ell_{max}$. Its value is also strongly
dependent on the frequency. In fact, for example in the case of no
clustering, the bias is directly proportional to the amplitude of the
shot--noise bispectrum of point sources (see
Eq.\,\ref{eq:fnl_bias_point_sources}).

If we consider only temperature and $S_c=0.3$\,Jy, the bias is
practically negligible at 165 and 225\,GHz (i.e. where
$b^{ps,white}_{\ell_1\ell_2\ell_3}$ is minimum) for the local,
orthogonal and flat shapes. The relative bias is less than 1 also for
the equilateral shape, which is the shape most affected by point
sources. As for $\sigma(f_{nl})$, the bias strongly increases at
$375\,$GHz, being typically more than one order of magnitude higher
than at 225\,GHz. At 105\,GHz it is still small, with a relative
value less than 1 (apart from the equilateral shape, with $\Delta
f_{nl}=107$). At lower frequencies, however, $\Delta
f_{nl}\gg\sigma(f_{nl})$.

Moreover, we find that $\Delta f_{nl}$ for the local, equilateral and
flat shapes is negative, although negligible, at low multipoles ($\ell
\le 200$) and it becomes positive at higher multipoles\footnote{This
  change in the sign of the ``local'' bias is not found in
  \citet{lacasa2012}, where the bias due to IR sources is still
  negative at $\ell=700$ and 2048. This different result may be
  dependent on the actual details of the characterization of the IR
  sources.}. The bias for the orthogonal shape is oscillating around
zero and therefore, in this case, there are certain multipoles where
$\Delta f_{nl}\simeq0$. 

It is interesting to notice that when instrumental noise and beam
function are introduced, the bias (and the relative bias) are equal or
even lower than in the ideal case at $\ell_{max}=2048$. One simple
explanation is that the beam and the noise prevent the high--$\ell$
terms in the sum of Eq.\,\ref{eq:fnl_bias_point_sources} and
\ref{eq:fnl_bias_point_sources_pol} to diverge.

When the polarization is considered, the bias suffers important
variations for each different shape in its amplitude and its
sign. This is dependent on the extra terms in the estimator: the
E-mode primordial bispectrum, the E-mode point source bispectrum and
also the inverse of the covariance matrix in
Eq. \ref{eq:covariance_bispectrum_pol}.  The bias of the local and
flat shapes becomes negative, whereas in the equilateral and
orthogonal shapes it keeps a positive value. The local and flat shapes
bias are also reduced in amplitude but the strongest improvement is
for the equilateral shape. On the contrary, at certain frequencies the
amplitude of the bias of the orthogonal shape increases compared with
the case of temperature only (see Tables \ref{table_ideal_bias_0d3}
and \ref{table_ideal_bias_1d0}). Finally, it can be noticed that the
amplitude of the bias is typically lower in a COrE--like mission than
in the ideal case, as effect of considering instrumental properties in
Eq.\,\ref{eq:fnl_bias_point_sources} and
\ref{eq:fnl_bias_point_sources_pol}.
\begin{table*}
\begin{center}
\caption{The expected uncertainty $\sigma(f_{nl})$, bias $\Delta f_{nl}$
  and relative bias $\Delta f_{nl}/\sigma(f_{nl})$ at
  $\ell_{max}=2048$ from temperature only and temperature plus polarization
  for a COrE--like mission. $S_c=$ 0.3 Jy is adopted.
  \label{table_core_bias_0d3}}
\begin{tabular}{c|c||cccccc|c|cccccc|}
\hline
& case & T & T &  T & T & T & T & & T+E  & T+E  & T+E &  T+E & T+E & T+E \\
& Freq. & 45 & 75 & 105 & 165 & 225 & 375 & & 45 & 75 & 105 & 165 & 225 & 375\\
\hline
& $\sigma(f_{nl})$ &  9.7 &        5.9 &        4.7 &        4.2 &        5.3  &       29.8    &&   6.0 &        3.5 &        2.7 &        2.3 &        2.5  &       20.6 \\
local & $\Delta f_{nl}$ & 16.0 &        4.3 &        1.8 &        0.4 &        0.7  &       25.1    &&   1.9 &       -0.8 &       -0.7 &       -0.1 &       -0.2  &        8.2 \\
& $\Delta f_{nl}/\sigma(f_{nl})$ & 1.7 &        0.7 &        0.4 &        0.1 &        0.1  &        0.8    &&   0.3 &       -0.2 &       -0.3 &       -0.1 &       -0.1  &        0.4 \\
\hline
& $\sigma(f_{nl})$ &84.3 &       65.2 &       58.1 &       56.0 &       65.2  &      232.6   &&   41.1 &       27.5 &       23.5 &       21.0 &       22.2  &      147.2 \\
equilateral &$\Delta f_{nl}$ & 421.9 &      202.4 &      102.7 &       17.5 &       49.0  &      935.9    &&  66.2 &       28.9 &       13.5 &        1.9 &        4.6  &      343.8 \\
& $\Delta f_{nl}/\sigma(f_{nl})$ &5.0 &        3.1 &        1.8 &        0.3 &        0.8  &        4.0    &&   1.6 &        1.1 &        0.6 &        0.1 &        0.2  &        2.3 \\
\hline
& $\sigma(f_{nl})$ &49.7 &       33.8 &       28.4 &       26.9 &       32.0  &      139.1  &&    21.7 &       13.8 &       11.4 &       10.1 &       10.6  &       82.0 \\
orthogonal & $\Delta f_{nl}$ & 39.9 &       12.9 &        6.5 &       -0.1 &        9.7  &      367.8  &&    12.1 &       23.8 &       16.1 &        2.8 &        4.4  &      130.2 \\
& $\Delta f_{nl}/\sigma(f_{nl})$ &0.8 &        0.4 &        0.2 &       -0.0 &        0.3  &        2.6    &&   0.6 &        1.7 &        1.4 &        0.3 &        0.4  &        1.6 \\
\hline
& $\sigma(f_{nl})$ &93.8 &       61.2 &       50.9 &       48.1 &       57.7  &      285.1   &&   39.1 &       24.5 &       20.2 &       17.9 &       18.9  &      148.1 \\
flat & $\Delta f_{nl}$ & 190.3 &       68.1 &       29.0 &        6.5 &        3.4  &      -68.8  &&    10.4 &      -26.3 &      -20.4 &       -3.7 &       -5.2  &      -38.4 \\
& $\Delta f_{nl}/\sigma(f_{nl})$ &2.0 &        1.1 &        0.6 &        0.1 &        0.1  &       -0.2   &&    0.3 &       -1.1 &       -1.0 &       -0.2 &       -0.3  &       -0.3 \\
\hline
\end{tabular}
\caption{The expected uncertainty $\sigma(f_{nl})$, bias $\Delta f_{nl}$
  and relative bias $\Delta f_{nl}/\sigma(f_{nl})$ at
  $\ell_{max}=2048$ from temperature only and temperature plus polarization
  for COrE--like mission. $S_c=$ 1 Jy is adopted.
  \label{table_core_bias_1d0}}
\begin{tabular}{c|c||cccccc|c|cccccc|}
\hline
& case & T & T &  T & T & T & T & & T+E  & T+E  & T+E &  T+E & T+E & T+E \\
& Freq. & 45 & 75 & 105 & 165 & 225 & 375 & & 45 & 75 & 105 & 165 & 225 & 375\\
\hline
& $\sigma(f_{nl})$ & 11.1 &        6.4 &        5.0 &        4.4 &        5.3  &       29.9   &&    6.6 &        3.7 &        2.9 &        2.4 &        2.5  &       20.6 \\
local & $\Delta f_{nl}$ & 130.7 &       37.1 &       15.6 &        3.4 &        2.3  &       25.0   &&   15.3 &       -7.0 &       -6.0 &       -1.2 &       -0.6  &        8.1 \\
& $\Delta f_{nl}/\sigma(f_{nl})$ & 11.8 &        5.8 &        3.1 &        0.8 &        0.4  &        0.8   &&    2.3 &       -1.9 &       -2.1 &       -0.5 &       -0.2  &        0.4 \\
\hline
& $\sigma(f_{nl})$ & 92.4 &       68.9 &       60.6 &       57.1 &       65.7  &      230.0   &&   43.7 &       28.4 &       24.0 &       21.2 &       22.3  &      146.3 \\
equilateral & $\Delta f_{nl}$ & 3428.4 &     1731.9 &      916.2 &      175.0 &      135.3  &      933.6 &&    515.4 &      235.0 &      117.5 &       19.5 &       12.6  &      346.7 \\
& $\Delta f_{nl}/\sigma(f_{nl})$ &37.1 &       25.1 &       15.1 &        3.1 &        2.1  &        4.1    &&  11.8 &        8.3 &        4.9 &        0.9 &        0.6  &        2.4 \\
\hline
& $\sigma(f_{nl})$ &55.3 &       36.2 &       30.0 &       27.5 &       32.3  &      138.7   &&   23.3 &       14.3 &       11.7 &       10.2 &       10.7  &       81.8 \\
orthogonal & $\Delta f_{nl}$ & 553.1 &      161.5 &       79.9 &        4.4 &       16.4  &      374.0  &&   110.3 &      202.6 &      136.9 &       25.4 &       12.2  &      132.5 \\
& $\Delta f_{nl}/\sigma(f_{nl})$ &10.0 &        4.5 &        2.7 &        0.2 &        0.5  &        2.7   &&    4.7 &       14.2 &       11.7 &        2.5 &        1.1  &        1.6 \\
\hline
& $\sigma(f_{nl})$ & 105.8 &       65.8 &       53.9 &       49.3 &       58.3  &      284.9   &&   42.0 &       25.5 &       20.8 &       18.1 &       19.0  &      147.7 \\
flat &$\Delta f_{nl}$ & 1232.1 &      523.1 &      232.3 &       58.3 &       26.5  &      -72.1   &&   59.4 &     -227.2 &     -172.4 &      -33.0 &      -14.7  &      -39.3 \\
& $\Delta f_{nl}/\sigma(f_{nl})$ &11.6 &        7.9 &        4.3 &        1.2 &        0.5  &       -0.3   &&    1.4 &       -8.9 &       -8.3 &       -1.8 &       -0.8  &       -0.3 \\
\hline
\end{tabular}
\end{center}
\end{table*}
\section{Conclusions}
\label{sect:conclusions}
In this paper we present a full set of forecasts on the contamination
produced by unresolved extragalactic point sources on the primordial
non-Gaussianity $f_{nl}$ parameter -- hopefully detectable by the
analysis of future high--resolution CMB (all--sky and low--noise)
anisotropy maps -- in the frequency interval 45--375\,GHz. We have
considered two scenarios: a) an ideal case without instrumental noise
nor instrumental beams; b) a future experiment with COrE-like
instrumental parameters. We have characterized all source populations
relevant in this frequency interval
by using two recently published 
models: the
\citet{tucci2011} model for extragalactic radio sources and the
\citet{lapi2011} model for far--IR/sub--mm sources. It is important to
remind here that these two models have proved very successful in the
explanation of all of the most recent and published results on number
counts (and on other relevant statistics) of extragalactic point
sources in the frequency range from $\sim 30$ to $\sim 1000$
GHz. 

As a first step, the angular power spectrum and bispectrum due to
unresolved extragalactic point sources for both temperature and
polarization have been estimated and compared. According to the above
quoted models, the contaminating signal produced by far--IR/sub--mm
selected extragalactic sources starts to dominate over the ERS one at
$\nu\ga200$\,GHz in the temperature (or total intensity) angular
power spectrum and bispectrum. In spite of their very low (average)
level of polarized emission, this source population is also relevant
in polarization, especially at large scales, due to the enhanced
signal produced by the strong clustering of high--z far--IR/sub--m
spheroidal galaxies. On the contrary, their effect is completely
negligible at frequencies $\la100$\,GHz. On the other hand, unresolved
ERS constitute a more relevant contaminant in all the frequency range
considered here and, thus, their effect on the CMB anisotropy signal
has to be properly modeled, for a correct evaluation of their
impact. In any case, their signal is found to be dominant over the
intrinsic CMB one only at frequencies below 75 GHz, thus out of the
CMB cosmological window.

We have studied the expected impact of undetected point sources on the
$f_{nl}$ parameter in terms of uncertainty and bias, for the local,
equilateral, orthogonal and flat shapes. For a COrE--like mission, we
find out that the best frequency for a $f_{nl}$ analysis is $\sim$
165\,GHz. At this frequency (and $S_c=0.3\,$Jy), the effect of
unresolved point sources on the accuracy in the determination of
$f_{nl}$ is almost negligible. The uncertainty on $f_{nl}$ does not
significantly change whether or not their contribution is included, and
estimated values of $f_{nl}$ are essentially unbiased (apart from the
equilateral shape for which $\Delta f_{nl}/\sigma(f_{nl})\sim0.3$ at
$\ell_{max}=2048$). At the frequencies 75 and 105\,GHz, results are
similar to 165--GHz ones for the $\sigma(f_{nl})$, but $f_{nl}$ can be
significantly biased, mainly for the equilateral and the flat shape;
on the other hand, at 225\,GHz we observe an increase in the $f_{nl}$
error--bars. If we move far from this frequency range the capability
of an accurate $f_{nl}$ measurement fast degrades, especially if only
temperature is used.


When polarization is considered, error--bars in the $f_{nl}$ estimate
are reduced by at least a factor 2 for the ideal case \citep[in
  agreement with estimates in literature; see e.g.][]{babich2004} and
they are not affected by point sources in all the frequency range
75--225\,GHz. Moreover, polarization can also help to significantly
reduce the bias induced by unresolved sources. As example, for the
equilateral shape, i.e. the shape most affected by point sources, the
bias is a factor from 3 to 10 lower than from temperature only
analysis. We can say therefore that polarization allows to expand the
best range of frequencies where $f_{nl}$ can be safely determined.

Discrepancies in the far-IR source bispectrum are observed when
compared with \citet{lacasa2012}. They are attributed to the different
ways to model the far--IR sources (see Section \ref{sect:bias}). This
dependency might affect the bias for the considered shapes especially
at high frequencies (see Section \ref{sect:results}).

Finally, as expected, the bias induced by point sources is strongly
related to the flux density limit down to which sources are detected
and removed/masked. Our results seem to point out that $S_c=0.3\,$Jy
is low enough to have unbiased $f_{nl}$ values at least at 165 and
225\,GHz for a COrE--like mission. At different frequencies, lower
flux density limits are required. On the other hand, $f_{nl}$
uncertainty is weakly affected by the choice of $S_c$ because it is
dominated by the intrinsic $f_{nl}$ variance (i.e. independent of
contaminants), that decreases when increasing $\ell_{max}$. We want to
stress the importance of high--resolution (i.e., FWHM$<$10\,arcmin)
and polarization--sensitive instruments, in order to have significant
improvements in the precision of primordial $f_{nl}$ in future
experiments beyond Planck.
\section*{Acknowledgments}
The authors acknowledge partial financial support from the Spanish
Ministerio de Econom\'ia y Competitividad projects
AYA2010-21766-C03-01, AYA-2012-39475-C02-01 and the Consolider
Ingenio-2010 Programme project CSD2010-00064. JGN acknowledges
financial support from the Spanish National Research Council (CSIC)
for a JAE-DOC fellowship. MLC thanks the Spanish Ministerio de
Econom\'ia y Competitividad for a Juan de la Cierva fellowship. AC
acknowledges the CSIC and the Spanish Ministerio de Educaci\'on,
Cultura y Deporte for a postdoctoral fellowship at the Cavendish
Laboratory of the University of Cambridge (UK). The authors
acknowledge the computer resources, technical expertise and assistance
provided by the Spanish Supercomputing Network (RES) nodes at
Universidad de Cantabria and Universidad Polit\'ecnica de Madrid.  We
also acknowledge the use of LAMBDA, support for which is provided by
the NASA Office of Space Science. We have also used the software
package HEALPix \citep{healpix}.
%
%
%

%
\end{document}